\documentclass[a4paper]{article}
\usepackage{graphicx}
\usepackage[applemac]{inputenc}
\usepackage[T1]{fontenc}
\usepackage{amsmath,amssymb,amsfonts,stmaryrd}	
\usepackage[a4paper,left=2cm,right=2cm,top=2.6cm,bottom=2.6cm]{geometry}

\usepackage[final]{pdfpages}

\renewcommand{\vec}[1]{\mathbf{#1}}
\renewcommand{\d}{\mathrm{d}}

\usepackage{color}

\usepackage{cite}
\usepackage{caption}
\usepackage{setspace}
\usepackage{textcase}
\usepackage{titlesec}
\titleformat{\section}{\center \small \bfseries}{\thesection. }{0pt}{\MakeTextUppercase}   \titlespacing*{\section}{0pt}{6ex plus 1ex minus .2ex}{4ex plus .2ex}
\titleformat{\subsection}{\center \small \bfseries}{\thesubsection. }{0pt}{}    \titlespacing*{\subsection} {0pt}{3.25ex plus 1ex minus .2ex}{3ex plus .2ex}
\titleformat{\subsubsection}{\center \small \itshape}{\thesubsubsection. }{0pt}{}    \titlespacing*{\subsection} {0pt}{3.25ex plus 1ex minus .2ex}{2ex plus .2ex}
\titleformat{\paragraph}[runin]{\normalfont \itshape}{}{0pt}{}    \titlespacing{\paragraph}{\parindent}{*1}{3\wordsep}
\renewcommand{\thesection}{\Roman{section}}
\renewcommand{\thesubsection}{\Alph{subsection}}
\renewcommand{\thesubsubsection}{\arabic{subsubsection}}
\usepackage{indentfirst}

\let\oldthebibliography=\thebibliography
\let\oldendthebibliography=\endthebibliography
\renewenvironment{thebibliography}[1]{\oldthebibliography{#1}\setcounter{enumiv}{26}}{\oldendthebibliography}


\begin{document}
\setstretch{1.2}

\title{Supplementary methods}
\date{\empty}



# Emergence of macroscopic directed motion in populations of motile colloids

Antoine Bricard[1], Jean-Baptiste Caussin[1,2], Nicolas Desreumaux[1], Olivier Dauchot[3] & Denis Bartolo[1,2]

[1]*PMMH, CNRS UMR7636, ESPCI-ParisTech, Université Paris Diderot and Université Pierre et Marie Curie, 10, rue Vauquelin 75005 Paris France.*

[2]*Laboratoire de Physique, Ecole Normale Supérieure de Lyon, CNRS UMR5672, 46, allée d'Italie F69007 Lyon, France.*

[3]*EC2M, CNRS UMR7083 Gulliver, ESPCI-ParisTech, 10, rue Vauquelin 75005 Paris France.*

**From the formation of animal flocks to the emergence of coordinate motion in bacterial swarms, at all scales, populations of motile organisms display coherent collective motion. This consistent behavior strongly contrasts with the difference in communication abilities between the individuals. Guided by this universal feature, physicists have proposed that solely alignment rules at the individual level could account for the emergence of unidirectional motion at the group level[1–4]. This hypothesis has been supported by agent-based simulations[1,5,6]. However, more complex collective behaviors have been systematically found in experiments including the formation of vortices[7–9], fluctuating swarms[7,10], clustering[11,12], and swirling[13–16]. All these (living and man-made) model systems (bacteria[9,10,16], biofilaments and molecular motors[7,8,13], shaken grains[14,15] and reactive colloids[11,12]) predominantly rely on actual collisions to display collective motion. As a result, the potential local alignment**

1

**rules are entangled with more complex, and often unknown, interactions. The large-scale behaviour of the populations therefore strongly depends on these uncontrolled microscopic couplings that are extremely challenging to measure and describe theoretically.**

**Here, we demonstrate a new phase of active matter. We reveal that dilute populations of millions of colloidal rollers self-organize to achieve coherent motion along a unique direction, with very few density and velocity fluctuations. Identifying, quantitatively, the microscopic interactions between the rollers allows a theoretical description of this polar-liquid state. Comparison of the theory with experiment suggests that hydrodynamic interactions promote the emergence of collective motion either in the form of a single macroscopic flock at low densities, or in that of a homogenous polar phase at higher densities. Furthermore, hydrodynamics protects the polar-liquid state from the giant density fluctuations, which were hitherto considered as the hallmark of populations of self-propelled particles[2,3,17]. Our experiments demonstrate that genuine physical interactions at the individual level are sufficient to set homogeneous active populations into stable directed motion.**

Our system consists of large populations of colloids capable of self-propulsion and of sensing the orientation of their neighbors solely by means of physical mechanisms. We take advantage of an overlooked electrohydrodynamic phenomenon referred to as the Quincke rotation[18,19] (Fig. 1a). When an electric field $\mathbf{E_0}$ is applied to an insulating sphere immersed in a conducting fluid, above a critical field amplitude $E_Q$, the charge distribution at the sphere's surface is unstable to infinitesimal fluctuations. This spontaneous symmetry breaking results in a net electrostatic torque, which causes the sphere to rotate at a constant speed around a random direction transverse to $\mathbf{E_0}$[18]. We

2

exploit this instability to engineer self-propelled colloidal rollers. We use PMMA beads of radius $a = 2.4\,\mu$m diluted in an hexadecane solution filling the gap between two conducting glass slides. Once the particles have sedimented on the bottom electrode, we apply an homogeneous electric field, and indeed observe their high-speed rolling motion (Fig. 1a). Isolated rollers propel along random directions (Fig. 1b). Their velocity $v_0$ is set by $E_0$ and scales as $[(E_0/E_Q)^2 - 1]^{\frac{1}{2}}$ (Fig.1c and Supplementary Materials).

In view of questioning the emergence of collective unidirectional motion, we electrically confine the roller populations in racetracks periodic in the curvilinear coordinate $s$. Their width is $500\,\mu$m $< W <$ $5$ mm (Fig. 2a and Methods). During a typical 10-minute-long experiment, millions of rollers travel over distances as large as $10^5 - 10^6$ particle radii, which makes it possible to investigate exceptionally large-scale dynamics. At low area fraction, $\phi_0$, the rollers form an isotropic gaseous phase. They all move at the same velocity along random directions as would an isolated particle (Fig. 2b and Supplementary video S1). Increasing $\phi_0$ above a critical value $\phi_c$, we observe a clear transition to collective motion. A macroscopic fraction of the rollers self-organizes and cruises coherently along the same direction (Figs. 2c, 2d and Supplementary Videos S2, S3, S4). More quantitatively, we define a polarization order parameter $\Pi_0$ as the modulus of the time and ensemble average of the particle-velocity orientation. $\Pi_0$ increases sharply with $\phi_0$ and displays a clear slope discontinuity at $\phi_c = 3 \times 10^{-3}$, revealing the strongly collective nature of the transition (Fig. 2e). Remarkably, $\phi_c$ is a material constant: it is independent of the electric-field amplitude.

3

For area fractions higher than but close to $\phi_c$, small density excitations nucleate from an unstable isotropic state and propagate along random directions. After complex collisions and coalescence events, the system phase separates to form a single macroscopic band that propagates at a constant velocity $c_{\text{band}}$ through an isotropic gaseous phase (Figs. 2a, 2c, 3a and Supplementary Videos S2 and S3). No stationary state involving more than a single band was observed even in the largest systems (10 cm long). The velocity $c_{\text{band}}$ is found to be very close to the single particle velocity $v_0$ at the front of the band. The bands are coupled to a net particle flux: they are colloidal flocks traveling through an isotropic phase. Their density profile is strongly asymmetric, unlike the slender bands observed in dense motility assays[7]. This marked asymmetry is akin to that found in 1D agent-based models[20]. It might be promoted by the high aspect ratio of the confinement. The local area fraction $\phi(s,t)$, at curvilinear coordinate $s$, and time $t$, increases sharply and then decays exponentially to a constant value $\phi_\infty$, which is very close to the critical volume fraction $\phi_c$ (Fig. 3b). This shape is similar to the one found in numerical simulations of the celebrated 2D Vicsek model[21,22]. Remarkably, the bands have no intrinsic scale and their length, $L_{\text{band}}$, is set by the particle-number conservation only. This result is readily inferred from Fig. 3c, which shows that the bands span a fraction of the racetrack which merely increases with $\phi_0$ regardless of the overall curvilinear length $L$.

Looking now at the local polarization, we observe that the colloidal flock loses its internal coherence away from the band front as $\Pi(s,t)$ decays continuously to zero along the band. Quantitatively $\phi(s,t)$ and $\Pi(s,t)$ are related in a universal manner irrespective of the particle velocity, and of the mean volume fraction (Fig. 3d). All our data collapse on a single master curve solely

parametrized by the particle fraction $\phi_\infty$ away from the band: $\Pi(s,t) = \left(1 - \frac{\phi_\infty}{\phi(s,t)}\right)$. As it turns out, this relation corresponds to particle-number conservation in a system where density and polarization waves propagate steadily at a velocity $v_0$ (see[22] and Supplementary Methods). This robust observation unambiguously demonstrates that the band state corresponds to a genuine *stationary* flocking phase of colloidal active matter.

Further increasing the area fraction above $\phi_0 \sim 2 \times 10^{-2}$, transient bands eventually catch-up with themselves along the periodic direction and form a homogeneous polar phase (Fig. 2d and Supplementary Videos S4) where the velocity distribution condenses on a single orientation of motion (Fig. 4a, to be contrasted with the perfectly isotropic distribution below $\phi_c$ in Fig. 1b). Conversely the roller positions are weakly correlated as evidenced by the shape of the pair-distribution function, which is similar to that found in low-density molecular liquids (Fig. 4b). We also emphasize that the density fluctuations are normal at all scales (Fig. 4c). This is the first experimental observation of a polar-liquid phase of active matter. The existence of a polar-liquid phase was theoretically established yet had never been observed in any prior experiment involving active materials. Until now, collective motion has been found to occur in the form of patterns with marked density and/or orientational heterogeneities[7,10,13,14,16]. Furthermore, in contrast with the present observations, giant density fluctuations are considered to be a generic feature of the uniaxially-ordered states of self-propelled-particle liquids[2,3,17]. We solve this apparent contradiction below and explain quantitatively our experimental observations.

From a theoretical perspective, the main advantage offered by the rollers is that their interac-

5

tions are clearly identified. We show in the Supplementary Methods how to establish the equations of motion of Quincke rollers interacting via electrostatic and far-field hydrodynamic interactions. They take a compact form both for the position $\mathbf{r_i}$, and the orientation $\hat{\mathbf{p}}_i$ of the $i^{\text{th}}$ particle:

$$\dot{\mathbf{r}}_i = v_0 \hat{\mathbf{p}}_i \qquad (1)$$

$$\dot{\theta}_i = \tau^{-1} \sum_{i \neq j} \partial_{\theta_i} \mathcal{H}_{\text{eff}}(\mathbf{r}_i - \mathbf{r}_j, \hat{\mathbf{p}}_i, \hat{\mathbf{p}}_j) \qquad (2)$$

where $\hat{\mathbf{p}}_i$ makes an angle $\theta_i$ with the $x$-axis. In dilute systems, the particle interactions do not affect their propulsion speed, yet the electric field and flow field compete to align the $\hat{\mathbf{p}}_i$s with them. This competition takes the form of an effective potential $\mathcal{H}_{\text{eff}}$ for the local orientations $\hat{\mathbf{p}}_i$. At leading order in $a/r$, $\mathcal{H}_{\text{eff}}(\mathbf{r}, \mathbf{p}_i, \mathbf{p}_j) = A(r)\hat{\mathbf{p}}_i \cdot \hat{\mathbf{p}}_j + B(r)\hat{\mathbf{p}}_i \cdot \hat{\mathbf{r}} + C(r)\hat{\mathbf{p}}_i \cdot (2\hat{\mathbf{r}}\hat{\mathbf{r}} - \mathbf{I}) \cdot \hat{\mathbf{p}}_j$, where $A(r)$ is a positive function, and thus promotes the alignment of the neighboring rollers. Importantly, $A$ is dominated by a hydrodynamic interaction, which arises from a hydrodynamic-rotlet singularity screened over distances of the order of the chamber height[23]. $B(r)$ is also short ranged and accounts for a dipolar repulsion. Conversely, $C(r)$ is long ranged and decays algebraically as $r^{-2}$ due to unscreened potential-dipole flows induced by the roller motion in confinement[24]. Neither $B$ nor $C$ yield any net alignment interaction. Neglecting these two terms, our model built from the actual microscopic interactions would amount to the so-called flying $xy$-model introduced phenomenologically in[25]. We emphasize that $\mathcal{H}_{\text{eff}}$ is independent of $v_0$ and $E_0$, and that it is *not* specific to the Quincke mechanism. Its expression could have been deduced from generic arguments based on global rotational invariance. We then use a conventional kinetic-theory framework à la Boltzmann to derive the large-scale equation of motion for the density, and the polarization fields[22,25]. In the present case, this approximation was fully supported by the weak positional correlations in all

the three phases, as exemplified in Fig. 4b. The resulting hydrodynamic equations are shown in the Supplementary Methods. At the onset of collective motion, the magnitude of the terms arising from the long-range hydrodynamic interactions are negligible. We are therefore left with equations for $\phi$ and $\Pi$ akin to the Toner and Tu model [2,3]. However, we provide explicitly the functional form of the transport coefficients phenomenologically introduced in [2]. Accordingly, we find that the competition between the polar ordering (induced by the short-range hydrodynamic interactions) and rotational diffusion yields a mean-field phase transition between an isotropic and a macroscopically ordered state (see Supplementary Methods). The phase transition occurs above a critical fraction $\phi_c$ that does not depend on the particle velocity (i.e. on $E_0$), in agreement with our experiments: collective motion chiefly stems from hydrodynamic interactions between the electrically powered rollers. However, at the onset of collective motion (i.e. for $\phi_0 > \phi_c$), the homogeneous polar state is linearly unstable to spatial heterogeneities. Moreover, for $\phi_0 > \phi_c$, the compression modes are unstable eigenmodes of the isotropic state, in agreement with the emergence of density bands observed in the experiments, all starting from an homogeneous state and an isotropic velocity distribution.

We also rigorously establish a kinetic theory for the strongly polarized state reached for $\phi_0 \gg \phi_c$ (Supplementary Methods). In this regime, the short range electrostatic repulsion matters, causing the density fluctuations to relax and stabilizes the polar-liquid state. In addition, the long-range hydrodynamic interactions further stabilize the system by damping the splay modes of $\Pi$ [24], thereby suppressing the giant density fluctuations[26], in agreement with our unanticipated experimental findings (Fig. 4c and Supplementary Methods). We stress here that theses long-range

7

hydrodynamic interactions do not depend at all on the propulsion mechanism at the individual level. They solely arise from from the confinement of the fluid in the $z$-direction [24]. They are therefore *not* specific to the Quincke propulsion mechanism. The only way to destroy the robust polar-liquid phase is to forbid it geometrically by eliminating the angular periodicity of the confinement in the curvilinear coordinate. In rectangular geometries with large enough aspect ratio, we observe that the bands never relax but rather bounce endlessly against the confining box (Supplementary video S5). In confinement with an aspect ratio of order one, the band state is replaced by a single macroscopic vortex (Supplementary video S6).

We have engineered large-scale populations of self-propelled particles from which collective macroscopic polar motion emerges from hydrodynamic interactions at exceptionally small densities. In addition, we believe that control over their interactions, and the ease with which they can be confined in custom geometries, will extend the current paradigm of active matter to collective motion in more complex environments relevant to biological, robotic and social systems.

**METHODS**

We use commercial PMMA colloids (Thermo scientific G0500, 2.4$\mu$m radius), dispersed in a 0.15 mol.L$^{-1}$ AOT/hexadecane solution. The suspension is injected in a wide microfluidic chamber made of double-sided scotch tapes. The tape is sandwiched between two ITO-coated glass slides (Solems, ITOSOL30, 80 nm thick). To achieve electric confinement an additional layer of scotch tape including a hole having the desired geometry is added to the upper ITO-coated slide. The holes are made with a precision plotting cutter (Graphtec robo CE 5000). The gap between the

8

two ITO electrode is constant over the entire chamber $H = 220\mu$m. The electric field is applied by means of a voltage amplifier (Trek 606E-6). The colloids are observed with a Nikon AMZ1500 stereomicroscope (1X magnification) equipped with a dark-field illuminator, and with a Zeiss axiovert microscope (10X objective) for local measurements. In both case, high speed movies are taken with CMOS camera (Basler A CE) at frame rates comprised between 70 and 900 fps. The particles are detected to a 1 pixel accuracy by locating the intensity maxima on the experimental pictures. The particle trajectories are reconstructed using a conventional tracking code[27].

1. Vicsek, T., Czirók, A., Ben-Jacob, E., Cohen, I. & Shochet, O. Novel type of phase transition in a system of self-driven particles. *Phys. Rev. Lett.* **75**, 1226–1229 (1995).

2. Toner, J. & Tu, Y. & Ramaswamy, S. Hydrodynamics and phases of flocks. *Annals of Physics.* **318** 170–244 (2005).

3. Marchetti, M.C *et al*. Soft active matter. Preprint available at http://arxiv.org/abs/1207.2929 (2012).

4. Vicsek, T. & Zafeiris, A. Collective motion. *Phys. Rep.* **517**, 71–140 (2012).

5. Grégoire, G & Chaté, H. Onset of Collective and Cohesive Motion. *Phys. Rev. Lett.* **92**, 025702, (2004).

6. Buhl, J.&J. Sumpter, D. J. T. & Couzin,I.D. Hale. J.J. & Despland, E. & Miller, E. R. & Simpson, S. J. From Disorder to Order in Marching Locusts. *Science.* **312**, 1402–1406, (2006).

7. Schaller, V., Weber, C., Semmrich, C., Frey, E. & Bausch, A. R. Polar patterns of driven filaments. *Nature* **467**, 73–77 (2010).

8. Sumino, Y. *et al.* Large-scale vortex lattice emerging from collectively moving microtubules. *Nature* **483**, 448–452 (2012).

9. Wioland, H., Woodhouse, F. G., Dunkel, J., Kessler, J. O. & Goldstein, R. E. Confinement stabilizes a bacterial suspension into a spiral vortex. *Phys. Rev. Lett.* **110**, 268102 (2013).

10. Zang, H. P., Be'er, A., Florin, E.-L. & Swinney, H. L. Collective motion and density fluctuations in bacterial colonies. *Proc. Natl. Acad. Sci. USA* **107**, 13626–13630 (2010).

11. Theurkauff, I., Cottin-Bizonne, C., Palacci, J., Ybert, C. & Bocquet, L. Dynamic clustering in active colloidal suspensions with chemical signaling. *Phys. Rev. Lett.* **108**, 268303 (2012).

12. Palacci, J., Sacanna, S., Steinberg, A. P., Pine, D. J. & Chaikin, P. M. Living crystals of light-activated colloidal surfers. *Science* **339**, 936–940 (2013).

13. Sanchez, T., Chen, D. T. N., DeCamp, S., Heymann, M. & Dogic, Z. Spontaneous motion in hierarchically assembled active matter. *Nature* **491**, 431–435 (2012).

14. Deseigne, J., Dauchot, O. & Chaté, H. Collective motion of vibrated polar disks. *Phys. Rev. Lett.* **105**, 098001 (2010).

15. Kudrolli, A., Lumay, G., Volfson, D. & Tsimring, L. Swarming and swirling in self-propelled polar granular rods. *Phys. Rev. Lett.* **100**, 058001 (2008).

16. Dombrowski, C., Cisneros, L., Chatkaew, S., Goldstein, R. E., Kessler, J. O. *Phys. Rev. Lett.* **93**, 098103 (2004).

17. Aditi Simha, R. and Ramaswamy, S. Hydrodynamic Fluctuations and Instabilities in Ordered Suspensions of Self-Propelled Particles *Phys. Rev. Lett.*, **89**, 058101, 2002.

18. Quincke, G. Ueber Rotationen im constanten electrischen Felde. *Ann. Phys. Chem.* **59**, 417–486 (1896).

19. Melcher, J. R. & Taylor, G. I. Electrohydrodynamics: a review of the role of interfacial shear stresses. *Annu. Rev. Fluid. Mech.* **1**, 111–146 (1969).

20. O. J. OLoan, O.J. & Evans, M. R. Alternating steady state in one-dimensional flocking. *J. Phys. A* **32**, L99 (1999).

21. Chaté, H., Ginelli, F., Grégoire, G. & Raynaud, F. Collective motion of self-propelled particles interacting without cohesion. *Phys. Rev. E* **77**, 046113 (2008).

22. Bertin, E., Droz, M. & Grégoire, G. Hydrodynamic equations for self-propelled particles: microscopic derivation and stability analysis. *J. Phys. A* **42**, 445001 (2009).

23. Hackborn, W. W. Asymmetric Stokes flow between parallel planes due to a rotlet. *J. Fluid Mech.* **218**, 531 (2006).

24. Brotto, T., Caussin, J.-B., Lauga, E. & Bartolo, D. Hydrodynamics of confined active fluids. *Phys. Rev. Lett.* **110**, 038101 (2013).

25. Farrell, F. D. C., Marchetti, M. C., Marenduzzo, D. & Tailleur, J. Pattern formation in self-propelled particles with density-dependent motility. *Phys. Rev. Lett.* **108**, 248101 (2012).

26. Schaller, V. & Bausch, A. R. Topological defects and density fluctuations in collectively moving systems. *Proc. Natl. Acad. Sci. USA* **110**, 4488–4493 (2013).

27. Crocker, J. C. & Grier, G. Methods of digital video microscopy for colloidal studies. *J. Colloid Interface Sci.* **179**, 298–310 (1996).

**Author Contributions** A.B. and J.-B. C. have equal contributions. A.B. and N.D. performed the experiments. A. B., N. D., O. D. and D. B. analyzed the experimental results. D.B. conceived and designed the experiments. J.-B. C. and D. B. performed the theory and wrote the Supplementary Methods. J.-B. C., O. D. and D.B. wrote the paper.

**Author Information** Correspondence and requests for materials should be addressed to D.B. (email: denis.bartolo@ens-lyon.fr).

**Figure 1  Single-roller dynamics. a,** Sketch of the Quincke rotation and of the self-propulsion mechanisms, and superposition of ten successive snapshots of rolling colloids. Time interval: $5.6$ ms. Scale bar $50\,\mu$m. **b,** Probability distribution of the velocity vector ($v_\parallel$, $v_\perp$) for isolated rollers. e$v_\parallel$ corresponds to the projection of the velocity on the direction tangent to the racetrack-shaped confinement, see Fig.2. $v_\perp$ is normal to $v_{//}$. The probability distribution involves more than $1.4\times 10^5$ instantaneous-speed measurements. **c,** Modulus of the individual velocity $v_0$ with respect to the electrostatic field amplitude $E_0$. Inset : $v_0^2$ versus $E_0^{\,2}$. The black dots represent the maximum of the probability distribution, and the error bars are defined from the standard deviation.

**Figure 2  Transition to directed collective motion. a,** Dark field pictures of a roller population that spontaneously forms a macroscopic band propagating along the racetrack, $E_0/E_Q = 1.39$ and $\phi_0 = 10^{-2}$. Scale bar $5$ mm. **b, c, d** Close views on the roller populations. The arrows correspond to the roller displacement between two subsequent movie frames (180 fps).Three distinct regimes are observed when varying $\phi_0$. **b,** isotropic gas, $\phi_0 = 6\times 10^{-4}$, **c,** propagating band, $\phi_0 = 10^{-2}$, **d,** homogeneous polar liquid $\phi_0 = 1.8\times 10^{-1}$. Scale bar $500\,\mu$m. (Note that the apparent size of the particles is twice larger than $2a = 4.8\,\mu$m, due to light diffraction). **e,** The modulus of the average polarization, $\Pi_0$, increases sharply from zero when the average area fraction $\Phi_0$ exceeds $\Phi_c = 3\,10^{-3}$ and reaches $\Pi_0 \sim 1$ for the largest area fraction. $\Phi_c$ is independent of $E_0$. The error bars correspond to the standard deviations.

**Figure 3** **Propagating-band state. a,** Left: spatiotemporal diagram of the area fraction recorded along the curvilinear coordinate $s$. Right: temporal variations of the area fraction measured at the curvilinear coordinate $s = 0.8L$ (white dotted line on left panel), where $L$ is the overall length of the racetrack. **b,** Shape of the band for four different area fractions $\Phi_0 = 5.3\,10^{-3}, 7.8\,10^{-3}, 1.0\,10^{-2}$, and $1.5\,10^{-2}$. The local area fraction is plotted as a function of $s/L$. Inside the band $\phi(s)$ decreases exponentially towards $\Phi_\infty \simeq \Phi_c$. **c,** The band length rescaled by the stadium length $L_{\text{band}}/L$ increases with the area fraction $\Phi_0$ and is independent of the racetrack length ($L = 28$ mm, (white dots): $L = 50$ mm (grey dots), $L = 73$mm (black dots). The error bars represent the estimated error associated with the measurements $L_{\text{band}}$ from the plots shown in **b**. **d,** Modulus of the local polarization $\Pi(s)$ plotted versus $1 - \Phi(s)/\Phi_\infty$. The black dots correspond to averages over 5000 local measurements (grey dots). The error bars correspond to the standard deviations. The red curve is the theoretical prediction.

**Figure 4** **Polar-liquid state. a,** Probability distribution of the velocity vector $(v_{//}, v_\perp)$ in the the polar-liquid state, where $v_\parallel$ corresponds to the projection of the velocity on the direction tangent to the racetrack shown in Fig.2. $v_\perp$ is normal to $v_{//}$. The probability distribution involves more than $3.2 \times 10^7$ instantaneous-speed measurements. **b,** Pair correlation function of the particle position in the polar liquid state. **c,** Variance of the number of colloids $\Delta N^2$ scales linearly with the average number of colloids $N$ counted inside boxes of increasing size. $E_0/E_Q = 1.39$ and $\phi_0 = 9.5 \times 10^{-2}$.

14

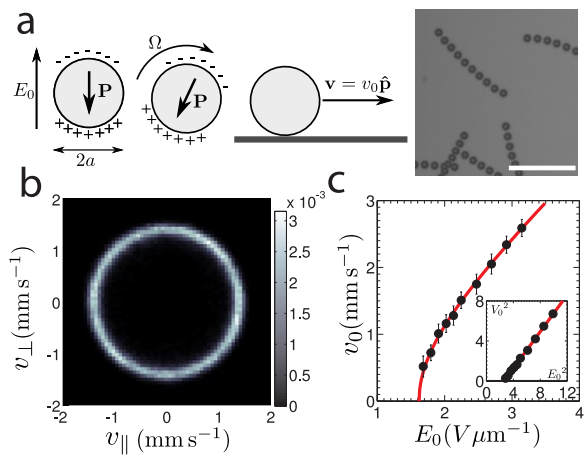

Bartolo - Figure 1

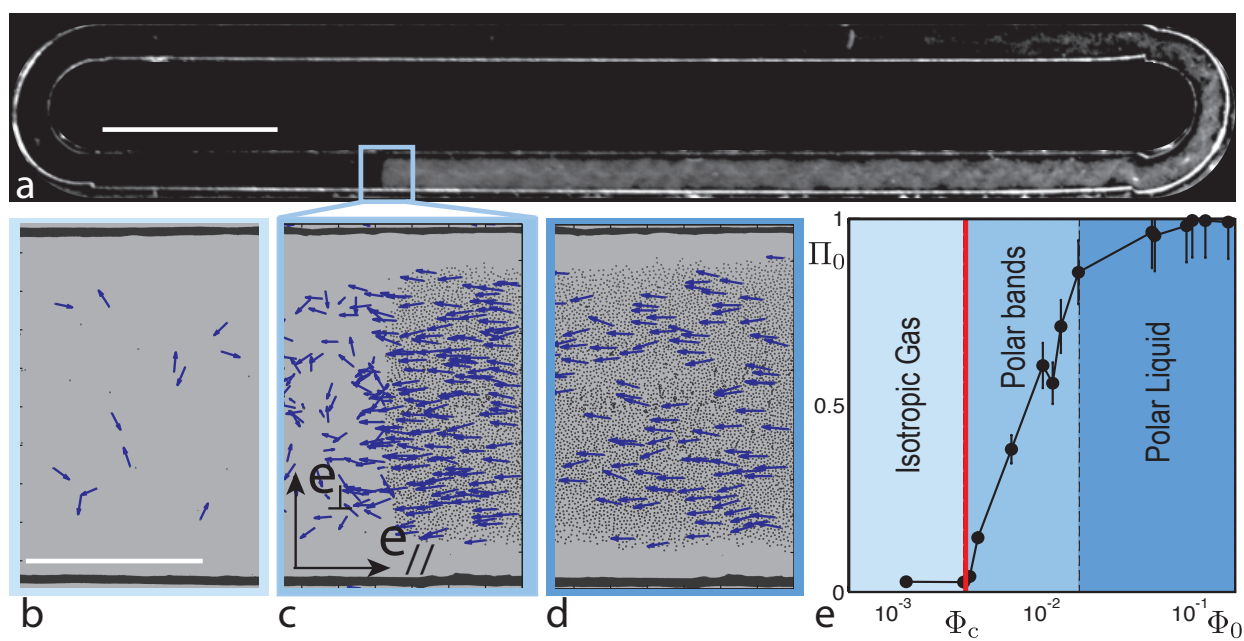

Bartolo - Figure 2

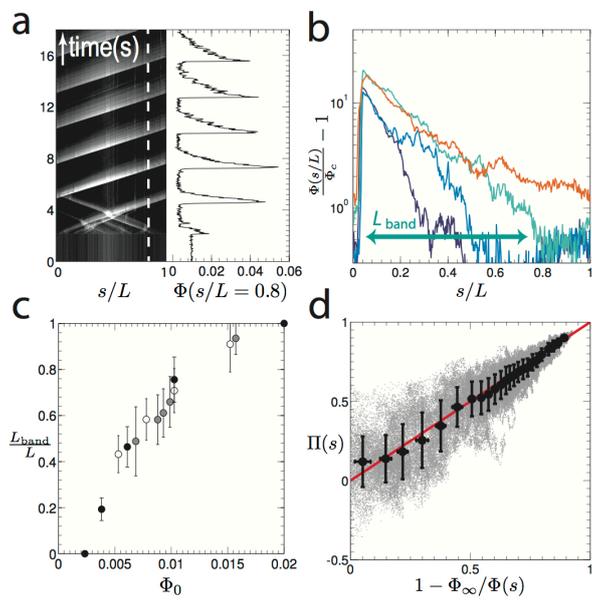

Bartolo - Figure 3

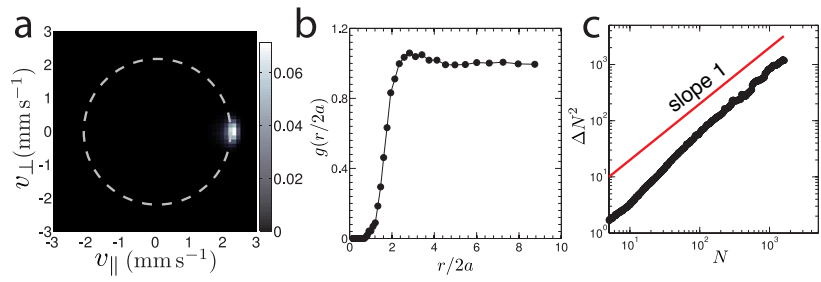

Bartolo - Figure 4



\maketitle

Here, we provide a comprehensive description of the theoretical model outlined in the main text, which accounts of the large-scale properties of a population of colloidal rollers.
For sake of clarity, this document is written in a self-consistent fashion, all the notations and definitions of  the main text are explicitly re-defined. It is organized as follows: In section~\ref{propulsion} we introduce a microscopic model that accounts for the motion of a single colloidal roller moving on a solid surface. Then, in section~\ref{interactions} we model the two-body interactions between colloidal rollers. We show that the combination of the electrostatic and the hydrodynamic couplings take the form of an effective potential $\cal H_{\rm eff}$ that couples the orientation of the rollers. In section~\ref{coarse-graining}, 
the latter microscopic model is coarse-grained following a kinetic theory framework. We  focus first on weakly polarized states, for which we establish  the dynamics of  the local density $\phi({\bf r},t)$, and of the local polarization field $\vec \Pi({\bf r},t)$ in section~\ref{transition}. This model accounts for a mean-field transition to collective motion. The linear stability of the homogeneous polar phase is questioned, and the existence of unstable compression modes is shown to be consistent with the formation of a band state. This stationary band state  is characterized by the constitutive relation between the local density and the local polarization.
Finally, we consider the large-scale dynamics of the polar-liquid phases in section~\ref{polar_liquid}. Our main result is the explanation for the suppression of the giant density fluctuations by long-ranged hydrodynamic interactions.

\section{From Quincke rotation to self-propulsion: the single roller dynamics}
\label{propulsion}

\subsection{Quincke rotation: uniform electric field, quiescent fluid}
\begin{figure}[!b] 
\begin{center}
\includegraphics[scale=0.34]{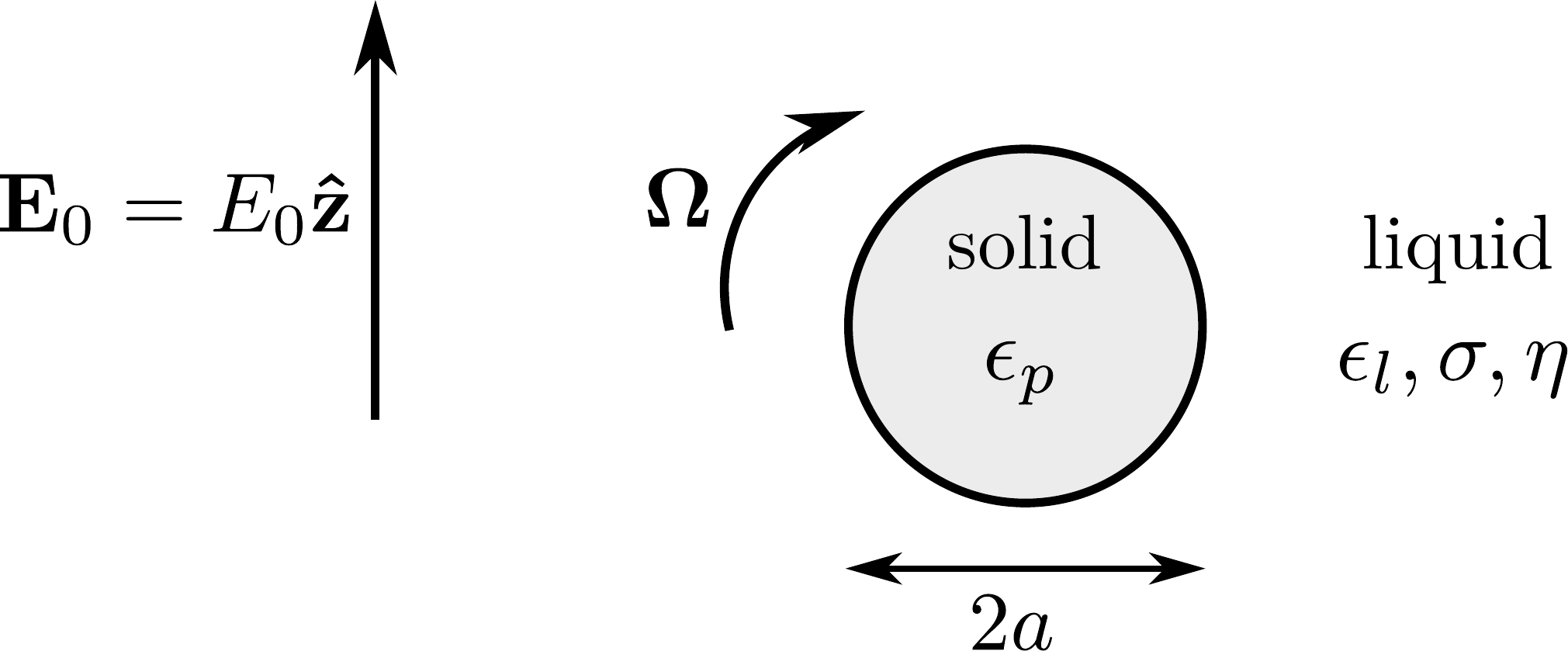} 
\end{center}
\caption{\setstretch{1.2}\label{fig1}An isolated solid sphere in an unbounded conducting liquid. When an external electric field $\vec E_0$ is applied, the particle can undergo Quincke rotation.}
\end{figure}

Before discussing the key role played by the solid surface, we briefly recall the main ingredients which originate the Quincke rotation of an isolated particle embedded in a quiescent and unbounded liquid~\cite{Melcher}. 

This electro-hydrodynamic effect arises from the interplay between interfacial electrodynamics and the particle motion in a viscous fluid.  Let us  consider an insulating sphere of radius $a$ located at $\vec r = \vec 0$, possibly rotating at the angular velocity $\vec \Omega$.  We note $\epsilon_p$ the dielectric permittivity of the particle. It is surrounded by a  conducting liquid with a conductivity $\sigma$ and a permittivity $\epsilon_l$. The solid particle is assumed to be impermeable. As the charge carriers in the liquid are ions, the sphere is  a perfect insulator. A uniform DC electric field $E_0 \, \vec{\hat z}$ is applied along the $z$-direction as sketched in Fig.~\ref{fig1}. After a transient regime, the electric charge relaxes to zero in the bulk.
 However, the charge distribution is not uniform at the the liquid-particle interface. Due to the conductivity and permittivity discontinuity, a non-uniform charge distribution arises close to the interface. The thickness of the charge layer is assumed to be much smaller than the particle radius $a$. Therefore it can be modeled by a  surface-charge distribution deduced from the continuity relation: $q_s = (\epsilon_l \vec E^l - \epsilon_p \vec E^p) \cdot \vec{\hat r} \vert_{r=a}$, where $\vec E^l$ (resp. $\vec E^p$) stands for the electrostatic field inside the liquid (resp. the particle).
Using the Maxwell's equations, it can be readily shown that the surface charge distribution is dipolar. It is thus described by its first moment $\vec P \equiv \int \d^2 s \; q_s \vec{\hat r}_s$. To establish how $\vec P$ depends on $\vec E_0$, we use the 
surface-charge conservation equation $\partial_t \, q_s + \nabla_s \cdot \vec j_s = 0$, where $\nabla_s \equiv (\vec I - \vec{\hat r} \vec{\hat r}) \cdot \nabla$ is the surface divergence operator, and $\vec j_s$ is the surface current. Due to the possible rotation of the particle, both ohmic conduction and charge advection contribute to the surface current: $\vec j_s = \sigma \vec E + q_s \, \vec \Omega \times a \vec{\hat r}$. After some elementary algebra, the charge-conservation equation can be recast  into a dynamical equation for the dipole moment $\vec P$~\cite{Melcher, Pannacci}:
\begin{equation}
\label{eqPtot}
        \frac{\d \vec P}{\d t} + \frac{1}{\tau} \vec P  = - \frac{1}{\tau}2 \pi \epsilon_0 a^3 \vec E_0 + \vec \Omega \times \left( \vec P - 4\pi \epsilon_0 a^3 \chi^\infty \vec E_0 \right)
\end{equation}
where $\chi^\infty \equiv \frac{\epsilon_p - \epsilon_l}{\epsilon_p+2\epsilon_l}$ and $\tau \equiv \frac{\epsilon_p + 2 \epsilon_l}{2 \sigma_l}$ is the so-called Maxwell-Wagner time.
It is convenient to distinguish  two contributions to the overall polarization vector: $\vec P\equiv\vec P^\epsilon+\vec P^\sigma$. The "static" contribution,  $\vec P^\epsilon \equiv 4 \pi \epsilon_0 a^3 \chi^\infty \vec E_0$ arises from the dielectric polarization, due to the permittivity discontinuity at the interface. The dynamic contribution $\vec P^{\sigma}$ results from the transport of the charges in the solution. When no rotation occurs, the dipole $\vec P$ relaxes towards a stationary value and orients along $- \vec E_0$ in a time $\tau$ due to the finite conductivity of the solution. However, as the particle rotates, charge advection competes with the spontaneous relaxation, and could in principle result in a dipole orientation making a finite angle with the external field $\vec E_0$. 

More quantitatively, we now show that the surface charge distribution can spontaneously break a rotational symmetry and therefore induce the steady rotation of the insulating sphere. In order to do so, we need an extra equation that is the angular momentum conservation. Since the particle carries surface charges, it may experience a net electric force $\vec F^{\rm e}$ and an electric torque $\vec T^{\rm e}$. The net interfacial electric stress is the jump of the Maxwell stress tensor across the interface:  $\vec{\hat r} \cdot \left[\boldsymbol{\mathcal T_M}^l - \boldsymbol{\mathcal T_M}^p \right]_{r=a}$, where $\boldsymbol{\mathcal T_M} \equiv \epsilon \vec E \vec E - \frac{1}{2} \epsilon E^2 \vec I$. Integrating over the surface, we obtain the torque $\vec T^{\rm e} = \frac{\epsilon_l}{\epsilon_0} \vec P \times \vec E_0$ and the net force $\vec F^{\rm e} = \frac{\epsilon_l}{\epsilon_0} (\vec P \cdot \nabla) \vec E_0$, which vanishes in a uniform external field.
Having colloidal systems in mind, we ignore the inertia of the sphere. Therefore the translation velocity $\vec v$ and the rotation speed $\vec \Omega$ are linearly related to $\vec F^{\rm e}$ and $\vec T^{\rm e}$ through a mobility matrix $\boldsymbol{\mathcal M}$:
\begin{equation}
\label{mobility_unbounded}
        \begin{pmatrix} \frac{1}{a} \vec v \\ \vec \Omega \end{pmatrix} \equiv \boldsymbol{\mathcal M} \cdot \begin{pmatrix}a \vec F^{\rm e} \\ \vec T^{\rm e} \end{pmatrix}
\end{equation}
In an unbounded fluid, $\boldsymbol{\mathcal M}$ is diagonal and has the form $\boldsymbol{\mathcal M} = \begin{pmatrix} \mu_t \vec I & 0 \\ 0 & \mu_r \vec I \end{pmatrix}$, where $\mu_t = (6 \pi \eta a^3)^{-1}$ and $\mu_r = (8 \pi \eta a^3)^{-1}$ in a liquid of viscosity $\eta$.
\begin{figure}
\begin{center}
\includegraphics[scale=0.36]{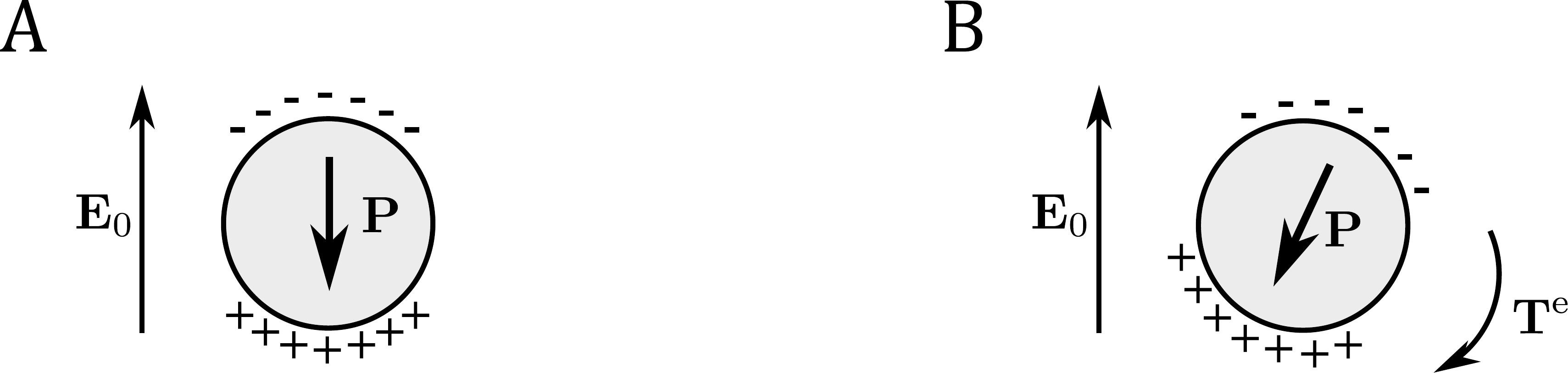} 
\end{center}
\caption{\setstretch{1.2}\label{Quincke}Emergence of Quincke rotation. {\bf A}--~Electric charges accumulate at the particle-liquid interface and result in a dipolar surface distribution. {\bf B}--~A small rotational perturbation tilts the dipole $\vec P$, thereby inducing a net electric torque $\vec T^{\rm e}$ which amplifies the initial perturbation.}
\end{figure}
Eqs.~(\ref{eqPtot}) and~(\ref{mobility_unbounded}) together fully capture the particle dynamics. When $\chi^\infty + \frac{1}{2} > 0$ and when the external field $\vec E_0$ exceeds the threshold value $E_{\rm Q} = \left[ 4 \pi \epsilon_l a^3 (\chi^\infty +\frac{1}{2}) \mu_r \tau \right]^{-1/2}$, two stationary states are found from Eqs.~(\ref{eqPtot}) and~(\ref{mobility_unbounded}). A first non-rotating solution is unstable against rotational perturbations. The second solution is stable and corresponds to a steady rotation at angular velocity
\begin{equation}
        \Omega = \frac{1}{\tau} \sqrt{\left(\frac{E_0}{E_{\rm Q}}\right)^2 -1}
\end{equation}
The rotation axis can be any direction perpendicular to $\vec E_0$ as the symmetry is spontaneously broken.
Such a stationary rotating state is conditioned by the competition between two opposite effects, that can be easily read from Eq.~(\ref{eqPtot}). On the one hand, the charge relaxation promotes the alignment of the dipole $\vec P$ in the direction $- \vec E_0$, thereby canceling the electric torque $\vec T^{\rm e}$. On the other hand, any small rotational perturbation advects the surface charge distribution and tilts the dipole $\vec P^\sigma$. This gives rise to a net electric torque $\vec T^{\rm e}$ which tends to amplify the initial disturbance, until it is balanced by the viscous torque, see Fig.~\ref{Quincke}. 

The Quincke-electro-rotation mechanism can be summarized as follows: when the external field exceeds a threshold value $E_{\rm Q}$, any infinitesimal perturbation results in an electrostatic torque which is large enough to advect the charges despite the stabilizing mechanism provided by the finite conductivity of the solution. The advection amplifies the initial perturbation until the viscous torque balances the electric torque.  When the stationary state is reached, the  particle  steadily rotates at $\Omega$ around an axis perpendicular to $\vec E_0$, the direction of which is set by the initial perturbation only.\\
\subsection{Self-propulsion of a Quincke roller}
In an unbounded fluid and a uniform electric field, the particle experiences no net force and thus have no translational velocity. To achieve  propulsion of the spheric particle, the basic idea is to let it roll on a plane surface that is one of the two electrodes used to induce $\vec E_0$. In order to establish the equations of motion of a Quincke roller, we have to modify both the mechanical and the electrostatic equations introduced above. 

\begin{figure}
\begin{center}
\includegraphics[scale=0.33]{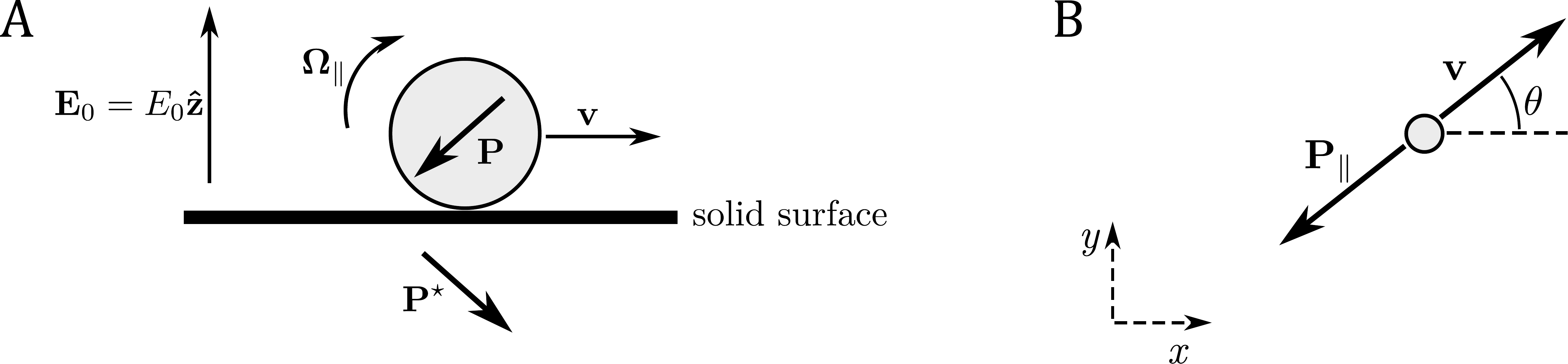}
\end{center}
\caption{\setstretch{1.2}\label{fig2}A Quincke rotating particle rolling on a plane conducting surface. {\bf A}--~The surface couples rotational and translational motion, allowing propulsion. It also disturbs the electric field. The dominant contribution to the image charge distribution is the symmetric dipole $\vec P^\star$. {\bf B}--~In the plane of the surface, the direction of the translation velocity is defined by the angle~$\theta$.}
\end{figure}

First we note that contact between the sphere and the planar electrode is lubricated by the surrounded liquid. A priori, the sphere both rolls and slides on the surface. This is accounted for by a modified mobility matrix $\boldsymbol{\mathcal M}$:
\begin{equation}
\label{mobility}
        \begin{pmatrix}  \frac{1}{a} \vec{v} \\ \vec{\Omega_\parallel} \\ \Omega_z        \end{pmatrix}   =   \boldsymbol{\mathcal M} \cdot     \begin{pmatrix}   a \vec F^{\rm e}_\parallel \\   \vec T^{\rm e}_\parallel \\     T^{\rm e}_z   \end{pmatrix}
\end{equation}
where we now have to distinguish between the in-plane components and the $z$-component of the vectors, see Fig.~\ref{fig2}A. 
The mobility matrix is non-diagonal and can be written as
\begin{equation}
\boldsymbol{\mathcal M} =  \begin{pmatrix}  \mu_t \vec I & \tilde \mu_t \boldsymbol{\Lambda}  & 0 \\ - \tilde \mu_r \boldsymbol \Lambda & \mu_r \vec I & 0 \\ 0 & 0 & \mu_\perp \end{pmatrix},
\end{equation}
where $\boldsymbol \Lambda = \begin{pmatrix} 0 & 1 & \\ -1 & 0 \end{pmatrix}$. The off-diagonal blocks of $\boldsymbol{\mathcal M}$ couple  rotational and translational motion: they are responsible for the rolling motion. 
For instance $\tilde \mu_t$ relies the electric torque to the translational propulsion speed. The mobility factors are easily inferred from the friction coefficients which were calculated in~\cite{Goldman1, Goldman2, O'Neil, Liu}, in the lubrication regime. They  depend only logarithmically on the distance between the particle and the surface, which was assumed to be small compared to the particle radius. Although we do not precisely control this gap, the numerical values of the mobility coefficients are weakly affected by this logarithmic dependence.

The surface at $z=0$ does not only modify the hydrodynamics of the  fluid, but also disturbs the electric field. Indeed, the particle lies on the lower electrode, which is an equipotential surface. We take it  into account by considering an electric image charge distribution in the region $z < 0$, which is dominated by a dipole $\vec P^\star = P_z \, \vec{\hat z} - \vec P_\parallel$ at $z = -a$, as sketched in Fig.~\ref{fig2}A. As opposed to the case considered in the previous section, where we considered the classical Quincke setup, the particle  experiences an external electric field which is {\em not} uniform. It includes here a correction $\delta \vec E^\star$ induced by the image charges. From the steady solution of the Quincke rotation in an infinite fluid, we can calculate the unperturbed dipole $\vec P$ in the absence of the surface, and then evaluate the disturbance field due to the surface. Whithin our experimental conditions, $E_0 <3 E_{\rm Q}$ and $\chi^\infty \ll \frac{1}{2}$~\cite{Pannacci}. Therefore,  the correction $\delta \vec E^\star$ is much smaller that the magnitude of the unperturbed field: $\vert \delta \vec E^\star \vert/E_0 \sim 0.01$. 
At leading order in $\vert \delta \vec E^\star \vert/E_0$, the dynamics of the electric polarization is written in a form that is  more complex than Eq.~(\ref{eqPtot}):
\begin{align}
\label{Pz}      &\frac{\d P^\sigma_z}{\d t} + \frac{1}{\tau} P^\sigma_z = \frac{\epsilon_l}{\epsilon_0} \mu_r E_0 {P^\sigma_\parallel}^2 - \frac{1}{\tau}4 \pi \epsilon_0 a^3\left(\chi^\infty + \frac{1}{2}\right) E_0\\
\label{Ppar}    &\frac{\d P^\sigma_\parallel}{\d t} + \frac{1}{\tau} P^\sigma_\parallel =  -\frac{\epsilon_l}{\epsilon_0} \mu_r E_0 P^\sigma_z P^\sigma_\parallel \\
\label{theta}   &\frac{\d \theta}{\d t} = 0
\end{align}
where $\theta$ defines the direction of the in-plane component of the polarization, Fig.~\ref{fig2}B.
The relation between the polarization, the electric torque and the electric force is not affected by the  substrate. In addition, it is worth noting that   the surface induces  no tangential force $\vec F^{\rm e}_\parallel$, and no perpendicular torque $T^{\rm e}_z = 0$ on the sphere. This is a rigorous results that does not depend on the specifics of the experiments. It holds at all order in $\vert \delta \vec E^\star \vert/E_0$ as it only follows from the symmetry of the 
real charges and of the image charges  with respect to the equipotential plane.  Combing now Eqs.~(\ref{eqPtot}) and~(\ref{mobility}) we infer the  equations of motion of an isolated sphere lying on a planar electrode: 
\begin{equation}
\label{eqPv}
        \vec v = - \frac{\epsilon_l}{\epsilon_0} a \tilde \mu_t E_0 \, \vec P^\sigma_\parallel
\end{equation}
This is the first main result of this supplementary document:  The particle steadily rolls on the electrode at a velocity $\vec v$, which points in the  direction opposite to the electric polarization. When $E_0 > E_{\rm Q}$,  the rolling speed $v_0\equiv|\vec v|$ is proportional to the in-plane component of $\vec P$, and is given by
\begin{equation}
        v_0 = \frac{a \tilde \mu_t}{\mu_r \tau} \sqrt{\left( \frac{E_0}{E_{\rm Q}} \right)^2 -1}
\end{equation}
The variations of the roller velocity that we measured are in excellent agreement with the above prediction as shown in  Fig.~1c main text. As our theory does not involve  any phenomenological parameter, we can provide an estimate of both the Quincke threshold and the  intrinsic velocity scale of the rollers.  We have $a = 2.4 \, {\rm \mu m}$, $\eta \sim 2 \, {\rm mPa} \cdot {\rm s}^{-1}$, $\epsilon_l \sim 2 \epsilon_0$ and $\tau \sim 1 \, {\rm ms}$~\cite{Pannacci}. So using the expressions below Eq. (\ref{mobility_unbounded}), we find $E_{\rm Q} \sim 10^6 \, {\rm V}\cdot {\rm m}^{-1}$ which is consistent with the value deduced from the best fit which yields  $E_{\rm Q} =1.6\,10^6 \, {\rm V}\cdot {\rm m}^{-1}$. The mobility coefficients weakly depends on  the thickness of the lubrication layer underneath the roller, which is assumed to be here of the order of 10-100 nm. Hence we find $\frac{a \tilde \mu_t}{\mu_r \tau} \sim 2\,{\rm mm\cdot s^{-1}}$, which again agrees well with the value deduced from our measurements $\sim 1.5\, {\rm mm\cdot s^{-1}}$. These results unambiguously confirm that the fast motion of the colloids results from the Quincke rotation of the colloids that in turn roll on the planar electrode, and that we have now a quantitative understanding of this novel self-propulsion mechanism.\\

\section{Roller-roller interactions}
\label{interactions}
The one-particle dynamics does not explicitly break the rotational invariance around the $\vec{\hat z}$-axis, as Eq.~(\ref{theta}) shows. When there are no inter-particle interactions, the system undergoes a spontaneous symmetry breaking, and all the rolling directions $\theta$ have the same probability. In this section we show how this  invariance is destroyed by the roller-roller interactions, and establish the equations of motion of a population of  interacting active colloids.
 
 The rollers are {\em a priori}  coupled by electrostatic and hydrodynamic interactions as well.  Their surface-charge distribution  induces a field disturbance $\delta \vec E (\vec r, t)$ which may alter the polarization, and the velocity, of the surrounding particles. Moreover, as it moves a roller induces a nontrivial fluid motion around it. Therefore, all the rollers are advected by a flow field $\vec u_\parallel(\vec r,t)$ resulting from the motion of their neighbors. Generically,  for a given distribution of roller position, both  $\delta \vec E_\parallel$ and  $\vec u_\parallel$  break the rotational invariance around $\vec{\hat z}$ thereby yielding orientational couplings between the active colloids.\\
 
\subsection{A roller in heterogeneous fields} 
Let us first consider the simpler problem of a single roller in a non-uniform  electric field. Its charge distribution has other multipolar components on top of the dipole that we considered so far. However, it can be shown that the dynamics of  $\vec P^\sigma$ is not coupled to the other multipoles, see e.g.~\cite{Das}, and obeys
\begin{equation}
\label{eqP}
        \frac{\d \vec P^\sigma}{\d t} + \frac{1}{\tau} \vec P^\sigma  = - \frac{1}{\tau}4 \pi \epsilon_0 a^3\left(\chi^\infty + \frac{1}{2}\right) \left( \vec E_0 + \delta \vec E \right) + \vec \Omega \times \vec P^\sigma
\end{equation}
This modified equation for the in-plane electric dipole is again complemented by a mechanical equation that relates the velocity of the particle to the forces and torques acting on it. Considering  that the roller is driven both by the Quincke mechanism and by an external fluid-flow field $\vec u$, we need to introduce two additional mobility coefficients to generalize Eq.~(\ref{mobility})~\cite{Goldman2}:
\begin{equation}
\label{mobility_shear}
        \begin{pmatrix}  \frac{1}{a} \vec{v} \\ \vec{\Omega_\parallel} \\ \Omega_z        \end{pmatrix}   =   \boldsymbol{\mathcal M} \cdot     \begin{pmatrix}   a \vec F^e_\parallel \\   \vec T^e_\parallel \\     T^e_z \end{pmatrix} + \begin{pmatrix}  \mu_s \left.\partial_z \vec u_\parallel \right\vert_{z=0}\\ \tilde \mu_s \left.\vec{\hat z} \times \partial_z \vec u_\parallel \right\vert_{z=0} \\0 \end{pmatrix}
\end{equation}
 
We now proceed to a perturbative analysis of Eqs.~(\ref{eqP}) and (\ref{mobility_shear}). Having dilute systems in mind, we assume that $\vert \delta \vec E \vert/E_0 = \mathcal O(\epsilon)$ and $\tau \left\vert \partial_z \vec u_\parallel  \right\vert = \mathcal O(\epsilon)$ are small parameters. We will further justify this approximation scheme at the end of this section. At leading order in $\epsilon$, Eqs.~(\ref{Ppar})--(\ref{eqPv}) remain valid despite the interactions. 
$P_z$, $P_\parallel$ and the norm of the velocity relax towards their unperturbed value over the timescale~$\tau$.
However, as anticipated the orientations of the particles are now coupled, and evolve on much longer timescales $\sim \tau/\epsilon$. After some tedious algebra, at leading order in $\epsilon$ this slow orientational dynamics  takes a rather simple form
\begin{equation}
\label{eq_theta}
        \frac{\d \theta}{\d t} = \frac{a}{\tau v_0} \frac{\tilde \mu_s \tilde \mu_t}{\mu_r} \, \vec{\hat p}^\perp \cdot \left.\partial_z \vec u_\parallel \right\vert_{z=0} - \frac{v_0}{a} \frac{\mu_r}{\tilde \mu_t} \left( \frac{\mu_\perp}{\mu_r} - 1 \right) \vec{\hat p}^\perp \cdot \frac{\delta \vec E_\parallel}{E_0} + \frac{a \tilde \mu_r}{\tau \mu_r}\, \vec{\hat p}^\perp \cdot (\vec P \cdot \nabla) \frac{\delta \vec E_\parallel}{E_0}
\end{equation}
where $\vec{\hat p}^\perp = -\sin \theta \, \vec{\hat x} + \cos \theta \, \vec{\hat y}$. The three terms on the right-hand-side have a clear physical meaning. First term: when the direction of motion is perpendicular to the flow field, the particle experiences a torque which promotes the reorientation along  the local flow direction. The direction $\vec{\hat p}$ thus rotates until aligning with $\vec u_\parallel$. Similarly, the second term accounts for the electrostatic coupling: it causes the dipole $\vec P^\sigma_\parallel$  to align with $\delta \vec E_\parallel$, and therefore aligns $\vec v$ in the opposite direction, since $(\frac{\mu_\perp}{\mu_r} - 1) > 0$. The last term stems from the field inhomogeneity. Within our experimental conditions, it can be checked from the numerical values of the prefactors, that this last term is subdominant, and we henceforth  neglect its contribution.

We have just shown that when a roller feels weak heterogeneities in the electric field and in the flow field, its propulsion  speed is unchanged. Conversely, the slow orientational dynamics of the particles  now explicitly breaks rotational invariance. The roller is prone to align its velocity with the reverse local electric field and with the local fluid velocity past the planar surface.\\
 
\subsection{Equation of motion of a population of interacting rollers} 
We now exploit the above results to establish the equations of motion of a population of interacting rollers. Eq.~(\ref{eq_theta}) is correct regardless of the origin of the fields' heterogeneities. Let us begin with a first important remark: If one now consider a test particle moving in an electric and a flow field perturbed by its neighbors, we readily infer that the speed of the test particle is unchanged.  In a dilute population of interacting rollers all the particles propel themselves at the same speed, which is again confirmed by the narrow velocity distributions found experimentally both in the isotropic and in the polar-liquid phases, Fig.~1b and 4a in the main text. 

To go beyond this  result, we  derive explicitly the forms of the electrostatic and of the hydrodynamic interactions between the active colloids.
\begin{figure}
\begin{center}
\includegraphics[scale=0.64]{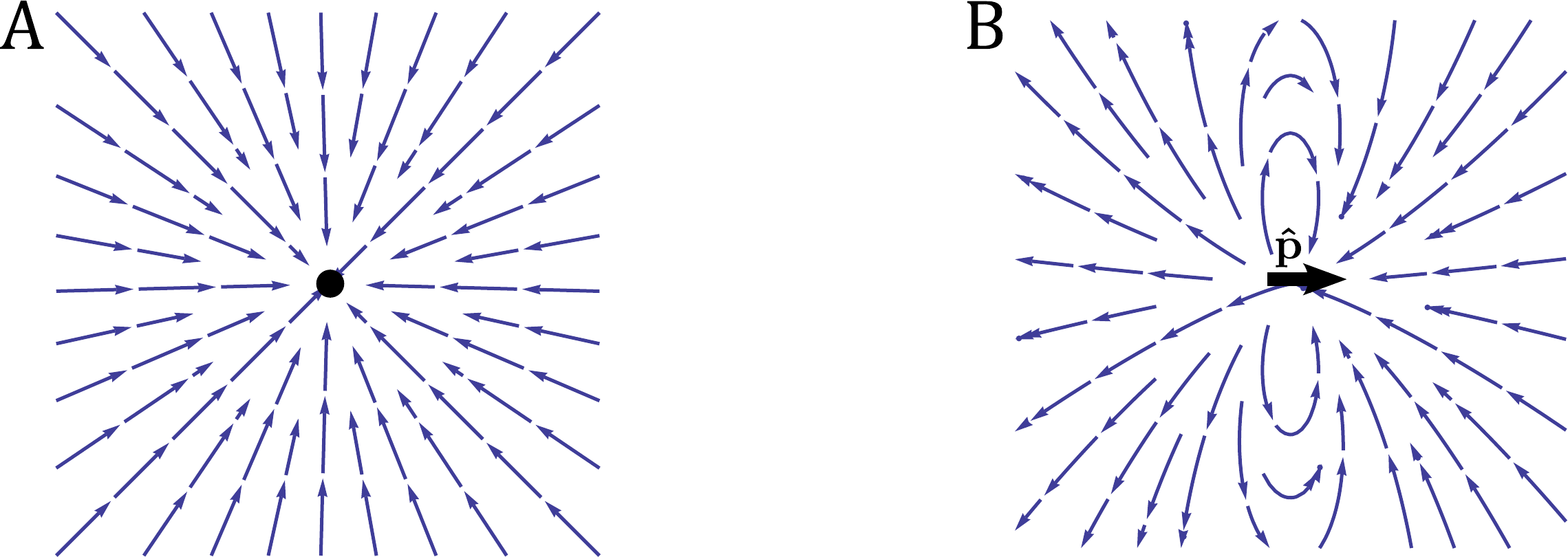} 
\end{center}
\caption{\setstretch{1.2}\label{elec}Electrostatic interactions: a particle rolling in direction $\vec{\hat p}$ creates a perturbative electric field. The field lines are plotted in the plane containing all the other particles, which tend to align in the opposite direction. {\bf A}--~A radial part (proportional to $P_z$) results in a repulsive effect, which does not depend on the orientation of the  particle. {\bf B}--~An additional contribution (proportional to $P_\parallel$) breaks the rotational symmetry and yields a position-dependent interaction.}
\end{figure}
We note $\vec r_i(t)$ (resp. $\vec{\hat p}_i(t)$) the position (resp. the orientation) of particle $i$.\\

\subsubsection{Electrostatic interactions}
We calculate the disturbance fields $\delta \vec E_\parallel(\vec r_i,t)$ and $\vec u_\parallel(\vec r_i,t)$ induced by all the other rollers $j \neq i$. The electric field induced by the particle $j$ originates from the dipole $\vec P_j$ and its electrostatic image $\vec P^\star_j$ (Fig.~\ref{fig2}A). Summing these two contributions in a far-field expansion, we find
\begin{equation}
\label{elec_field}
        \delta \vec E^{(j)}_\parallel(\vec r_i,t) = \frac{3}{2 \pi \epsilon_0 r_{ij}^3} \left[ \frac{a}{r_{ij}} P_z \, \vec{\hat r}_{ij} - \frac{a^2}{r_{ij}^2} P_\parallel \, \vec{\hat p}_j \cdot \left( 5 \vec{\hat r}_{ij} \vec{\hat r}_{ij} - \vec I \right) \mathcal + \mathcal O \left( \frac{a^3}{r_{ij}^3} \right) \right]
\end{equation}
where $\vec r_{ij} = \vec r_i - \vec r_j = r_{ij} \, \vec{\hat r}_{ij}$, and where $P_z$ and  $P_\parallel$ are the components of the total polarization at order $\epsilon^0$. We recall that in an heterogeneous field, the dipolar fouling to the local field causes the roller to align its velocity in a direction opposite to $\delta \vec E_{\parallel}$. Hence we infer from Eq.~(\ref{elec_field}) that the two-body electrostatic interactions combine two contributions. The first term in Eq.~(\ref{elec_field}) is proportional to $P_z$. Since $P_z < 0$, this first term corresponds to a repulsive interaction: it favors a roller velocity $\vec v_i$ pointing in the direction opposite to $\vec{\hat r}_{ij}$
The second term on the r.h.s of Eq.~(\ref{elec_field}) is proportional to $P_\parallel$, and it possibly results in alignment or anti-alignment with $\vec{\hat p}_j$, depending on the relative positions between the two rollers. The symmetry of these two electrostatic couplings is better understood by inspecting the  electric-vector field  plotted in Fig.~\ref{elec}. So far, we have implicitly neglected the influence of the upper electrode, which is also a conducting equipotential surface. The former results are therefore valid only at  distance  smaller than the separation distance $H$ between the two electrodes. Experimentally, the channel height is $H = 200 \, {\rm \mu m} \gg a$. At larger scales, all the electrostatic couplings are exponentially screened over a characteristic length $H/\pi$.

As a last comment about electrostatic interactions, we note that Eqs.~(\ref{elec_field}) and~(\ref{hydro_field}) confirm that the perturbative treatment $\vert \delta \vec E \vert/E_0 = \mathcal O(\epsilon)$, $\tau \left\vert \partial_z \vec u_\parallel  \right\vert = \mathcal O(\epsilon)$ is self-consistent for dilute systems  as the algebraic electrostatic repulsion   prevents the formation of concentrated clusters in a population of rollers.\\
\subsubsection{Hydrodynamic interactions}
\begin{figure}
\begin{center}
\includegraphics[scale=0.64]{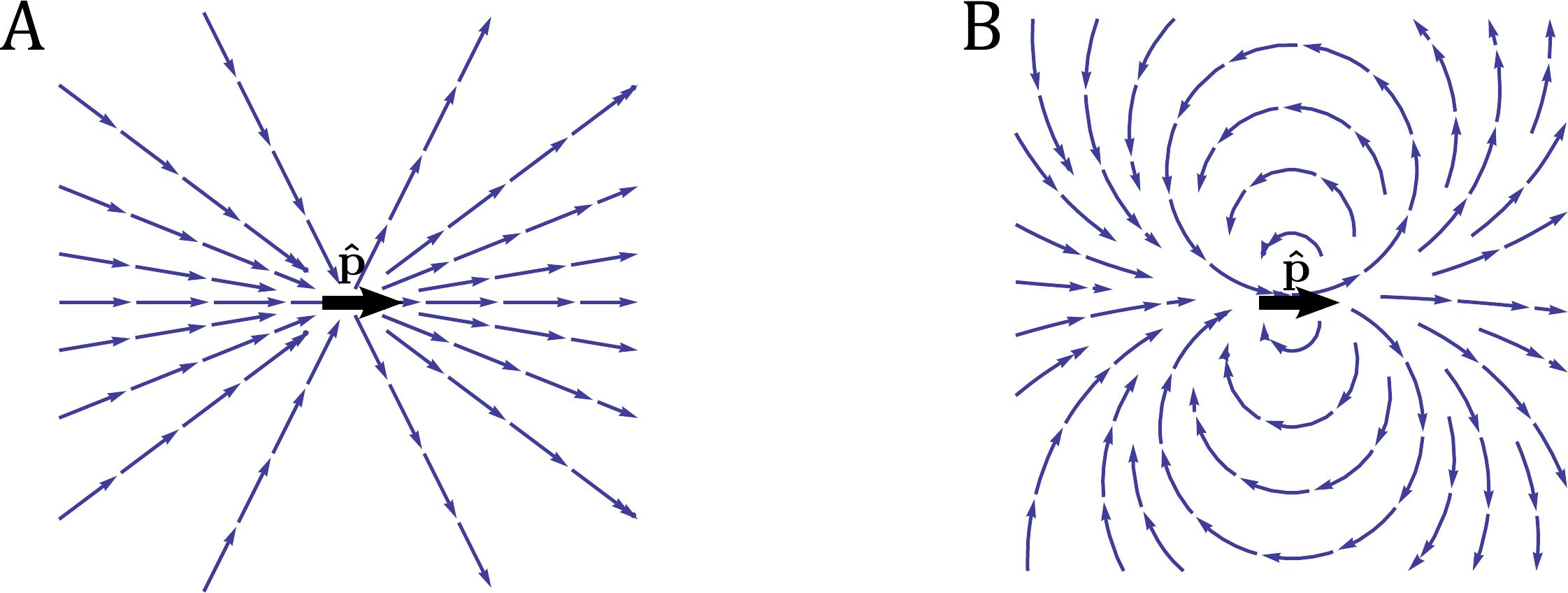} 
\end{center}
\caption{\setstretch{1.2}\label{hydro}Hydrodynamic interactions: a particle rolling in direction $\vec{\hat p}$ creates a flow field. The streamlines are plotted in the plane containing all the other particles, which tend to align in flow. {\bf A}--~At distances smaller than the channel height, the central roller induces a radial shear with anisotropic amplitude, which globally promotes alignment. {\bf B}--~At distances much larger than the channel height, the non-screened resulting flow has a dipolar symmetry.}
\end{figure}

A similar approach is used to deal with the hydrodynamic interactions in dilute systems. The flow field created by the particles is expressed in terms of pointwise hydrodynamic singularities. 
$r_{ij}<H$: Over distances smaller than the channel height $H$, a Quincke roller is akin to a rotlet near a no-slip wall. The particle is a pointwise torque-source which induces a  complex flow field. This flow  was computed by Blake and Chwang using the image singularity method~\cite{Blake}. $r_{ij}>H$: At long distances, unlike electrostatic screening, mass conservation gives rise to a non-vanishing flow having the form of a two-dimensional source dipole, as it was derived by Hackborn for a rotlet  located between two rigid walls~\cite{Hackborn}. These  results provide the shear rate induced by particle $j$ at the location of the particle $i$. Keeping only the leading order terms in a $a/r_{ij}$ expansion, we obtain
\begin{equation}
\label{hydro_field}
        \left.\partial_z \vec u^{(j)}_\parallel \right\vert_{z=0}(\vec r_i,t) = \begin{cases} \frac{6 \mu_r}{a \tilde \mu_t} v_0 \frac{a^3}{r_{ij}^3} (\vec{\hat p}_j \cdot \vec{\hat r}_{ij}) \vec{\hat r}_{ij}&\textrm{if }r_{ij} < \frac{H}{\pi}\\
        \frac{6 \mu_r}{H \tilde \mu_t} v_0 \frac{a^2}{r_{ij}^2} \, \vec{\hat p}_j \cdot (2 \vec{\hat r}_{ij} \vec{\hat r}_{ij} - \vec I)  &\textrm{at long distance }r_{ij} \gg \frac{H}{\pi} \end{cases}
\end{equation}
The corresponding streamlines are plotted in Fig.~\ref{hydro}. We showed that the particle velocity reorients along the local  direction of $\partial_z \vec u_\parallel$. Therefore, we deduce from Eq.~(\ref{hydro_field}) and Fig.~\ref{hydro}A that, at short distances ($r_{ij}\ll H$) the hydrodynamic interactions promote the alignment of the roller velocities. In addition,  for $r_{ij}>H$, long-range hydrodynamic interactions that algebraically decay as $r^{-2}$ have a dipolar symmetry. They can either cause alignment or anti-alignment, depending on the relative positions between the rollers, Fig.~\ref{hydro}B.\\
%
\subsubsection{Equations of motion}

Assuming that both electrostatic and hydrodynamic interactions are pairwise additive, the above results can be summarized in a compact form. The particle $i$ moves at constant velocity $v_0$ on the surface, and undergoes a slow orientational dynamics:
\begin{align}
\label{eq_motion1}	&\vec{\dot r}_i = v_0 \vec{\hat p_i}\\
\label{eq_motion2}	&\dot \theta_i = \frac{1}{\tau} \frac{\partial}{\partial \theta_i} \sum_{j \neq i} \mathcal H_{\rm eff} (\vec r_i - \vec r_j, \vec{\hat p}_i,  \vec{\hat p}_j) + \sqrt{2 D_r} \, \xi_i(t)
\end{align}
The global interaction potential $\mathcal H_{\rm eff}$ accounts for all  the possible interactions between the rollers that we have established above. It takes the generic  form:
\begin{equation}
\label{potential}
        \mathcal H_{\rm eff} (\vec r, \vec{\hat p}_i,  \vec{\hat p}_j) = A(r) \, \vec{\hat p}_j \cdot \vec{\hat p}_i + B(r) \, \vec{\hat r} \cdot \vec{\hat p}_i + C(r) \, \vec{\hat p}_j \cdot (2\vec{\hat r}\vec{\hat r} - \vec I  )\cdot \vec{\hat p}_i
\end{equation}
where the coefficients have complex expressions, deduced from well identified microscopic parameters:
\begin{align}
\label{A}       &A(r) = 3 \tilde \mu_s \frac{a^3}{r^3} \, \Theta\left(r\right) + 9 \left( \frac{\mu_\perp}{\mu_r} -1 \right) \left( \chi^\infty + \frac{1}{2} \right) \left( 1 - \frac{E_{\rm Q}^2}{E_0^2} \right) \frac{a^5} {r^5} \, \Theta\left(r\right)\\
        &B(r) =  6 \left( \frac{\mu_\perp}{\mu_r} -1 \right) \sqrt{\frac{E_0^2}{E_{\rm Q}^2}-1} \left[\left( \chi^\infty + \frac{1}{2}\right) \frac{E_{\rm Q}^2}{E_0^2} - \chi^\infty\right] \frac{a^4}{r^4} \, \Theta\left(r\right)\\
\label{C}        &C(r) = 6 \tilde \mu_s \frac{a}{H}\frac{a^2}{r^2} + 3 \tilde \mu_s \frac{a^3}{r^3} \, \Theta\left(r\right) + 15 \left( \frac{\mu_\perp}{\mu_r} -1 \right) \left( \chi^\infty + \frac{1}{2} \right) \left( 1 - \frac{E_{\rm Q}^2}{E_0^2} \right) \frac{a^5} {r^5} \, \Theta\left(r\right)
\end{align}
Here, $\vec r$ reduces to a two-dimensional vector parallel to the surface, and  $\Theta$ accounts for the screening of finite-range interactions. For  sake of simplicity, we henceforth approximate the screening function by a step function: $\Theta(r) = 1$ if $r \leq H/\pi$, and $\Theta(r) = 0$ otherwise. We have also introduced a noise term in Eq.~(\ref{eq_motion2}) to account for rotational diffusion. $\xi_i(t)$ is a Gaussian white noise with zero mean and unit variance $\langle \xi_i(t) \xi_j(t') \rangle = \delta (t-t') \delta_{ij}$. Remarkably, the rotational diffusivity  $D_r$ is the only phenomenological coefficient of our theory. 

Several comments are in order:

(i) The term $A(r) \, \vec{\hat p}_j \cdot \vec{\hat p}_i$ is an alignment interaction. It arises  both from the short-distance hydrodynamic interactions and from part of the electrostatic couplings. They correspond respectively to the first and the second terms in~(\ref{A}).

 (ii) In the absence of the $B$ and $C$ terms, our model would reduce to the so-called "flying XY model" introduced phenomenologically in~\cite{Farrell}. Nevertheless, additional terms have been obtained from the microscopic analysis.
 
  (iii) The coefficient $B(r)$ is  positive, since $\chi^\infty + \frac{1}{2}>0$ and $\chi^\infty < 0$ in our experimental system. It corresponds to the electrostatic repulsive coupling. The last term $C(r)$ combines electric and hydrodynamic interactions.
Contrary to $A(r)$, theses additional terms in Eq.~(\ref{potential}) do not yield any net alignment interaction in an isotropic population.

(iv) $A(r)$ and $B(r)$ are finite-range interactions, being screened on a distance set by the channel height. Conversely, $C(r)$ contains the unscreened dipolar hydrodynamic coupling. It is truly long-ranged since it algebraically decays like $r^{-2}$ in two dimensions. Note however, that its strength is small compared to the short-range hydrodynamic effect, since it is proportional to $a/H \ll 1$.

(v) The hydrodynamic interactions yield no dependence on $E_0$ for the effective potential~$\mathcal H_{\rm eff}$, as the first terms in Eqs.~(\ref{A}) and~(\ref{C}) show. Although the induced flow field is proportional to the particle velocity, the norm of the velocity vector $v_0 \vec{\hat p}$ is constant. As a consequence, the resulting orientation rate does not depend on $v_0$,  it is thus independent of the external electric field.

(vi) Finally we note, as we did it in the main text, that the generic relation~(\ref{potential}) is not specific to the Quincke rollers. 
 Indeed, this effective potential is expected whenever particles move at constant velocity, and experience short-range polar alignment. The slow angular variations of the two-body interactions  are described by the first terms of a generic Fourier-expansion in~$\theta$. Imposing global rotational invariance, the resulting effective potential can be recast into a function of $\vec{\hat p}$, leading to the generic equation~(\ref{potential}). In this approach, the first term accounts for alignment in a uniform field. The repulsive term proportional to $B(r)$ corresponds to a local alignment in a monopolar field, while the last term corresponds to a local alignment in a dipolar field. We stress that no other lower-order moment is allowed, due to symmetry considerations.  Within this framework, the flow induced by model swimmers referred to as pushers and pullers would be coupled via a higher order quadrupolar term reflecting the symmetry of the flow lines induced by force dipoles [47].\\

\section{From microscopic interactions to macroscopic hydrodynamic equations}
\label{coarse-graining}

In the following, we link the microscopic interaction rules to the large-scale properties of the roller population. The microscopic equations of motion have to be coarse-grained, in order to derive kinetic equations for hydrodynamic fields such as the particle density and the orientation field. We sumarize here the main steps of this procedure. Using standard kinetic theory methods (see e.g. \cite{Risken,Menzel}), the $2N$ coupled Langevin equations~(\ref{eq_motion1})--(\ref{eq_motion2}) can be transformed into a Fokker-Planck equation for the $N$-particle distribution function $\Psi^{(N)}(\vec r_1..., \vec r_N, \theta_1, ..., \theta_N, t)$:
\begin{equation}
	\frac{\partial \Psi^{(N)}}{\partial t} + \sum_i \nabla_i \cdot \left( v_0 \vec{\hat p}_i \Psi^{(N)} \right) + \sum_i \frac{\partial}{\partial \theta_i} \left( \frac{1}{\tau} \sum_{j \neq i} \frac{\partial \mathcal H_{\rm eff} (\vec r_i - \vec r_j, \theta_i,  \theta_j)}{\partial \theta_i} \Psi^{(N)} \right) - D_r \sum_i \frac{\partial^2}{\partial \theta_i^2} \Psi^{(N)} =0
\end{equation}
By integrating over $N-1$ particle positions and directions, we obtain the time evolution of the one-particle density $\Psi^{(1)}(\vec r, \theta, t) \equiv \frac{1}{(N-1)!} \int \d^2\vec r_2...\d^2 \vec r_N \d \theta_2 ... \d \theta_N \; \Psi^{(N)}(\vec r, \vec r_2,... \vec r_N, \theta, \theta_2,...,\theta_N,t)$. It is coupled to the two-point distribution function $\Psi^{(2)}(\vec r,\vec r', \theta, \theta', t) \equiv \frac{1}{(N-2)!} \int \d^2\vec r_3...\d^2 \vec r_N \d \theta_3 ... \d \theta_N \; \Psi^{(N)}(\vec r, \vec r', \vec r_2,... \vec r_N, \theta, \theta', \theta_3,...,\theta_N,t)$, and obeys:
\begin{equation}
\label{Psi_1}
	\partial_t \Psi^{(1)} + v_0 \, \vec{\hat p} \cdot \nabla \Psi^{(1)} + \frac{1}{\tau} \partial_\theta  \int d^2 \vec r' \d \theta' \; \frac{\partial \mathcal H_{\rm eff} (\vec r - \vec r', \theta,  \theta')}{\partial \theta}  \Psi^{(2)}(\vec r, \vec r',  \theta, \theta', t) - D_r \, \partial_\theta^2 \Psi^{(1)} = 0
\end{equation}
The latter expression is the first equation of an infinite hierarchy, which couples the $n$-point  distribution $\Psi^{(n)}$ to the $(n+1)$-point distribution $\Psi^{(n+1)}$. To close this hierarchy of equations, we postulate a relation between $\Psi^{(2)}$ and $\Psi^{(1)}$, and introduce a generalized mean-field (i.e. Boltzmann-like) approximation. 
We assume that the two-body correlations cancel over a distance as small as one particle diameter. We also include steric exclusion effects between the colloids:
\begin{equation}
\label{closure1}
	\Psi^{(2)}(\vec r, \vec r', \theta, \theta', t) = \begin{cases} 0 &\textrm{if } \vert \vec r - \vec r' \vert < 2a \\
	\Psi^{(1)}(\vec r, \theta, t) \Psi^{(1)}(\vec r', \theta', t) &\textrm{if } \vert \vec r - \vec r' \vert \geq 2a \end{cases}
\end{equation}
This ansatz is  supported by the absence of positional correlation in the three phases (gas, bands and polar liquid). Even at the high densities, in the polar-liquid phase, the radial distribution function of the colloids is very well approximated by a Heaviside function. In addition, we note that this approximation was successfully used to describe the large scale behavior of driven confined suspensions~\cite{Desreumaux}.
We then derive from Eqs.~(\ref{Psi_1}) and (\ref{closure1}) a closed equation for the one-particle distribution function.
The hydrodynamic fields that characterize the structure of the population are defined by the angular  Fourier modes of $\Psi^{(1)}$. Defining these modes as $\Psi^{(1)}(\vec r, \theta, t) = \frac{1}{2 \pi} \sum_{k \in \mathbb Z} \hat \Psi^{(1)}_k(\vec r,t) \, \mathrm{e}^{-i k \theta}$, the three hydrodynamic field that we consider are:
\begin{alignat}{3}
	&\textrm{Area fraction:} \quad &&\phi(\vec r,t) \equiv \frac{1}{\pi a^2} \int \d \theta  \; \Psi^{(1)}(\vec r, \theta, t) =  \frac{1}{\pi a^2} \hat \Psi^{(1)}_0\\
	&\textrm{Velocity polarization:} \quad&&\vec{\Pi}(\vec r,t) \equiv \frac{\pi a^2}{\phi} \int \d \theta  \; \vec{\hat p} \Psi^{(1)}(\vec r, \theta, t) = \frac{1}{\hat \Psi^{(1)}_0} \begin{pmatrix} \mathrm{Re} \,\hat \Psi^{(1)}_1 \\ \mathrm{Im} \,\hat \Psi^{(1)}_1	\end{pmatrix}\\
	&\textrm{Nematic order tensor:} \quad&&\vec Q(\vec r,t) \equiv \frac{\pi a^2}{\phi} \int \d \theta  \; \left( \vec{\hat p} \vec{\hat p} - \frac{1}{2}\vec I \right) \Psi^{(1)}(\vec r, \theta, t) = \frac{1}{2 \hat \Psi^{(1)}_0} \begin{pmatrix}  \mathrm{Re} \, \hat \Psi^{(1)}_2 & \mathrm{Im} \, \hat \Psi^{(1)}_2 \\ \mathrm{Im} \, \hat \Psi^{(1)}_2 & - \mathrm{Re} \, \hat \Psi^{(1)}_2 \end{pmatrix}
\end{alignat}
In all that follows and in the main text we do not refer anymore to the electrostatic properties of the colloids. Therefore, for sake of simplicity $\vec \Pi$ will be simply referred to as the polarization field. 

By integrating Eq.~(\ref{Psi_1}) over $\theta$, we immediately recover the particle-number conservation law:
\begin{equation}
\label{continuity}
	\partial_t \, \phi + v_0 \nabla \cdot (\phi \vec \Pi) = 0
\end{equation}
Taking the first angular moment of Eq.~(\ref{Psi_1}) similarly couples the time evolution of $\vec \Pi$ to the nematic order tensor~$\vec Q$. We thereby generate a new hierarchy of equations which couples each moment of the distribution function to higher-order moments~\cite{Hinch, Baskaran, Woodhouse}. One more closure assumption is required, and it should be carefully defined for each  phase that we want to describe as we will show it below.\\

\section{Transition to collective motion}
\label{transition}
We first focus on the transition to collective motion. For weakly-polarized phases, two possible closure schemes have been used in the context of active fluids. Bertin \textit{et al.}~\cite{Bertin, Peshkov} introduced  a scaling ansatz for the amplitude of the angular Fourier components of the one-point function. This ansatz is expected to be relevant for nearly-isotropic states with small and slow variations of the hydrodynamic field. Baskaran and Marchetti~\cite{Baskaran} assume the distribution function to be a linear functional of its first three moments. This ansatz is obviously exact in the limit of purely isotropic states. Coming back to our model for the population of rollers,~(\ref{eq_motion1})--(\ref{potential}), we have checked that these two closure methods lead to the same kinetic equations, and are therefore strictly equivalent. We also assume that $\vec Q$ is a fast-relaxing variable, following again~\cite{Bertin} and \cite{Hinch} in a fluid mechanics context. Within this approximation scheme, after lengthy algebra, and at leading order in $a/H \ll 1$, we obtain the following equation for the evolution of the orientation field:
\begin{align}
\label{eq_isotropic}
	&\tau \partial_t (\phi \vec \Pi) + \frac{3 v_0 \alpha}{8 D_r} (\phi \vec \Pi \cdot \nabla) \, \phi \vec \Pi = \left[\alpha \, \phi - \tau D_r - \frac{\alpha^2}{2 \tau D_r} (\phi^2 \Pi^2) \right] \phi \vec \Pi + \kappa \phi \, \vec M * \phi \vec \Pi - \frac{1}{2}\left( \tau v_0 + a \beta \phi \right) \nabla \phi \\& \hspace{5cm} \nonumber  - \frac{5 v_0 \alpha}{8 D_r} (\nabla \cdot \phi \vec \Pi) \, \phi \vec \Pi   +\frac{5 v_0 \alpha}{16 D_r} \nabla (\phi^2 \Pi^2)+ \frac{\alpha \beta}{2\tau D_r} a (\nabla \phi \cdot \phi \vec \Pi) \, \phi \vec \Pi + \mathcal O(\nabla^2)
\end{align}
where
\begin{align}
\label{alpha}	&\alpha \equiv \int_{r \geq 2a} \!\!\!\! \d r \; A(r) \frac{r}{a^2} = \frac{3}{2} \tilde \mu_s + \frac{3}{8} \left( \frac{\mu_\perp}{\mu_r} -1 \right) \left( \chi^\infty + \frac{1}{2} \right) \left( 1 - \frac{E_{\rm Q}^2}{E_0^2}\right)\\
\label{beta}	&\beta \equiv \int_{r \geq 2a} \!\!\!\! \d r \; B(r) \frac{r^2}{a^3} = 3 \left( \frac{\mu_\perp}{\mu_r} -1 \right) \sqrt{\frac{E_0^2}{E_{\rm Q}^2}-1} \left[\left( \chi^\infty + \frac{1}{2}\right) \frac{E_{\rm Q}^2}{E_0^2} - \chi^\infty\right]\\
\label{kappa}	&\kappa \equiv 3 \tilde \mu_s \frac{a}{H} \ll \alpha\\
	&\vec M * \phi \vec \Pi(\vec r,t) \equiv \frac{1}{\pi} \int_{\vert \vec r - \vec r'\vert \geq 2a} \!\!\!\! \d^2 \vec r' \; \frac{1}{\vert \vec r - \vec r' \vert^2} \left(2 \frac{(\vec r - \vec r')(\vec r - \vec r')}{\vert \vec r - \vec r' \vert^2} - \vec I\right) \cdot \phi(\vec r',t) \vec \Pi(\vec r',t)
\end{align}
We stress that all the coefficients involved in the above non-local equation have been inferred  from a well controlled  microscopic model introduced in the first section of this document. We only briefly recall their physical origin:
\begin{itemize}
\item $\alpha > 0$~accounts for the alignment interactions, which favor the emergence of polar order.  It is chiefly set by the local hydrodynamic interactions between the rollers (first term on the r.h.s of Eq.~\ref{alpha}).  It yields the same generic terms as those found in~\cite{Bertin} or~\cite{Farrell} (when the particle velocity is constant), which are known to lead to large-scale coherent motion. 
\item $\beta > 0$ stems from the repulsive electrostatic couplings. 

\item $\kappa$ gives the strength of the long-range dipolar hydrodynamic interactions, which result in a non-local operator $\vec M$. We studied the impact of these truly long-range interactions in~\cite{Desreumaux, Brotto}.\\

\end{itemize}
 \subsection{Homogeneous states: A Curie-Weiss description of collective motion}
 \begin{figure}
\begin{center}
\includegraphics[width=\textwidth]{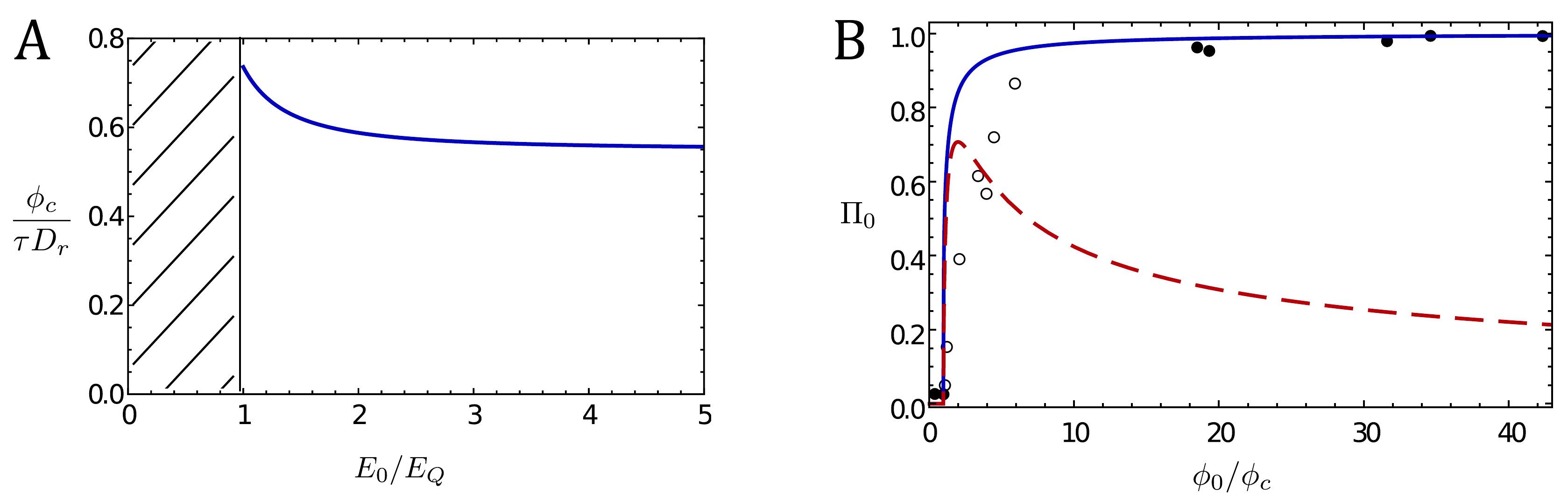} 
\end{center}
\caption{\setstretch{1.2}\label{fig_transition}Homogeneous steady states. {\bf A}--~Transition line in the plane of density and external applied electric field. The variations of the critical area fraction (rescaled by $\tau D_r$) are plotted as a function of $E_0/E_{\rm Q}$. (The distance between the particle and the surface was set to $0.01 a$, mobility coefficients were deduced from~\cite{Goldman1, Goldman2, O'Neil, Liu} and dielectric constants were taken from~\cite{Pannacci}). {\bf B}--~Orientation--density relation in homogeneous phases. Red dashed line: bifurcation curve given by Eq.~(\ref{bifurc}), valid when $\phi_0/\phi_c$ is close to $1$. Blue full line: prediction from Eq.~(\ref{order_polar}), accurate for strongly polar phases. The symbols denote the experimental values of the average polarization. The data show a transition to collective motion at a critical area fraction, as expected from both theoretical curves. They also unveil the formation of polar-liquid phases (filled symbols) at high area fractions, in agreement with Eq.~(\ref{order_polar}). However, for intermediate densities (open symbols), the experimental data correspond to phase-separated states consisting in  bands propagative in a gaseous apolar phase. The band state is obviously not accounted for by the theoretical expressions~(\ref{bifurc}) and~(\ref{order_polar}), which hold for spatially homogeneous systems only. As a consequence, the theoretical curves are quantitatively relevant for isotropic phases and polar-liquid phases only. However we stress that the two asymptotic models correctly predict a phase transition to a macroscopically polar state.}
\end{figure}
Looking for homogeneous phases, i.e. dropping space derivatives, Eq.~(\ref{continuity}) reduces to $\phi(\vec r,t)=\phi_0$, and  Eq.~(\ref{eq_isotropic}) takes the simple form:
 \begin{align}
 \label{CW_isotropic}
	&\tau \partial_t \vec \Pi = \left(\alpha \, \phi_0 - \tau D_r \right) \vec \Pi - \frac{\alpha^2}{2 \tau D_r} (\phi_0^2 \Pi^2) \vec \Pi 
\end{align}
 Hence, it readily follows from the cubic form of the r.h.s that the system undergoes a mean-field phase transition to a polar state as $\phi_0$ exceeds  the critical area fraction:
\begin{equation}
	\phi_c = \frac{\tau D_r}{\alpha}
\end{equation}
At small density $\phi_0 \leq \phi_c$, the only stationary state is an isotropic phase with zero mean orientation: $\Pi_0 = 0$. The disordered solution becomes unstable above $\phi_c$, and the system forms a polar ordered phase with $\Pi_0 \neq 0$. At the onset of collective motion, the following bifurcation is expected, see also Fig.~\ref{fig_transition}B:
 \begin{equation}
 \label{bifurc}
	\Pi_0 (\phi_0) = \begin{cases} \sqrt{2 \frac{\phi_c}{\phi_0} \left( 1 - \frac{\phi_c}{\phi_0} \right)} &\textrm{if } \phi_0 > \phi_c \\
	0  &\textrm{if } \phi_0 \leq \phi_c \end{cases}
\end{equation}
 Starting from a realistic and accurate  microscopic description of the Quincke mechanism at the single-roller level, we have established the existence of a genuine phase transition to collective motion in  populations of such active colloids. This is one of our main theoretical results.
 
 To further stress on  the importance of the hydrodynamic interactions  in this collective phenomena, we plot the variations of $\phi_c$ as  a function of $E_0$ in Fig.~\ref{fig_transition}A. Using microscopic parameters corresponding to our experimental setup, we indeed observe that the transition line weakly depends on the magnitude of external field.  As we discussed above, the hydrodynamic polar interactions result in an orientation rate which does not depend on the particle velocity, and thus does not yield any dependence on $E_0$ for $\phi_c$. 
The qualitative agreement with the experimental data, which  showed no dependence of $\phi_{\rm c}$ on $E_0$, strengthens  the relevance of our theory, further confirms the central role of the hydrodynamic interactions in the transition to collective motion.\\

 \subsection{Linear stability analysis}
\label{linearstabilityweaklypolar}
We now investigate the stability of the isotropic and of the polar homogeneous phases against spatial fluctuations. We consider plane-wave perturbations: $\phi(\vec r,t) = \phi_0 + \delta \hat \phi \, e^{i(\vec q \cdot \vec r - \omega t)}$ and $\vec \Pi(\vec r,t) = \Pi_0 \vec{\hat x} + \delta \vec {\hat \Pi} \, e^{i(\vec q \cdot \vec r - \omega t)}$. In the following, we define the wave-vector direction as $\vec q \equiv q \cos \varphi_\vec q \, \vec{\hat x}+ q \sin \varphi_\vec q \, \vec{\hat y}$.\\
 
\begin{figure}
\begin{center}
\includegraphics[scale=0.75]{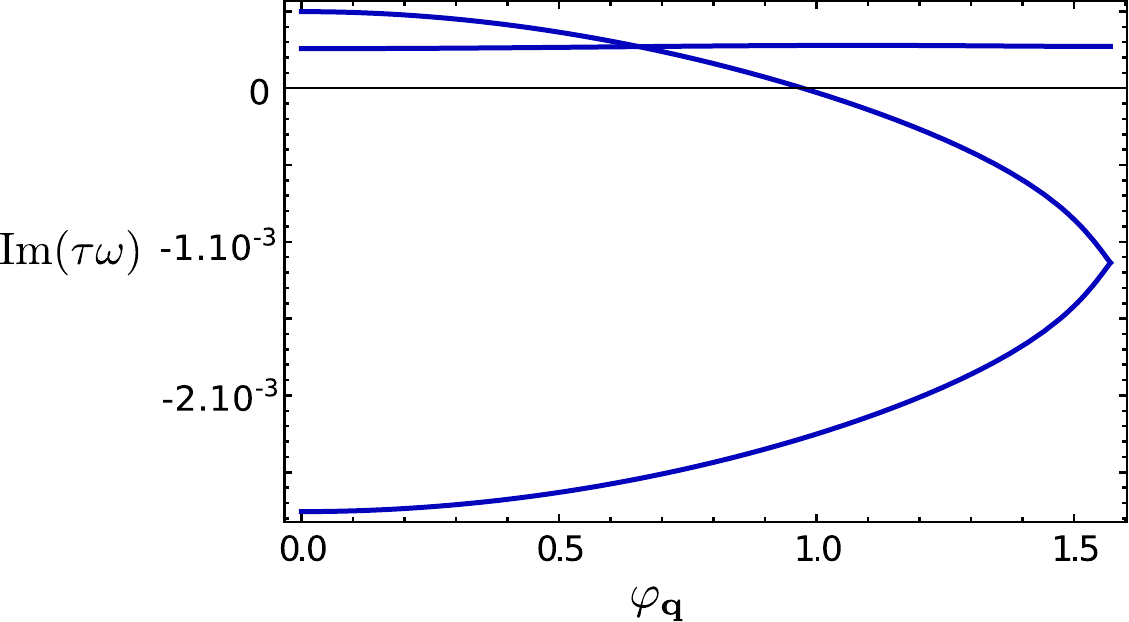} 
\end{center}
\caption{\setstretch{1.2}\label{stab1}Stability of weakly-polar states against linear fluctuations. The growth rates of the three eigenmodes modes are plotted as a function of the wave-vector direction $\varphi_\vec q$, for $E_0 = 2 E_Q$, $\phi_0 = 1.1 \phi_c$ and $qa = \frac{1}{500}$. The values of the other parameters are the same as in Fig.~\ref{fig_transition}. Instabilities were observed for all values of $E_0$, $\phi_0$, and $qa$.}
\end{figure}

 \paragraph*{Weakly-polarized phases} We linearize Eqs.~(\ref{continuity}) and~(\ref{eq_isotropic}) around a uniform polar state with density $\phi_0 > \phi_c$ and orientation $\Pi_0 = \sqrt{2 \frac{\phi_c}{\phi_0} \left( 1 - \frac{\phi_c}{\phi_0} \right)}$. The eigenvalues $\omega(\vec q)$ cancel a cubic equation which is solved numerically. We found that the growth rates ${\rm Im}(\omega)$ are invariant upon the transformations $\varphi_\vec q \rightarrow - \varphi_\vec q$ and $\varphi_\vec q \rightarrow \varphi_\vec q + \pi$. Without loss of generality, we focus on $\varphi_\vec q \in \left[ 0 , \frac{\pi}{2} \right]$. The growth rates are plotted in Fig.~\ref{stab1} as a function of the wave vector direction $\varphi_\vec q$. The uniform polar states are unstable for all wave-vector directions: no homogeneous polar phase exists at the onset of collective motion. This result is consistent with all the available models that were used so far to account for the emergence of collective dynamics in active polar matter~\cite{Marchetti}. The compression modes are eigenmodes of the linear stability problems. They are exponentially amplified in the polar phase. This mode of destabilization is consistent with the  formation of propagative bands in our experiments for $\phi_0$ close but above $\phi_{\rm c}$.
 \paragraph*{Isotropic phases}
 A similar stability analysis is carried out around a uniform and isotropic state $\Pi_0 = 0$. Two modes couple the density fluctuations and the  orientational perturbations in the longitudinal direction. The corresponding eigenvalues are $\omega_\pm= i \frac{\alpha}{2 \tau} (\phi_0 - \phi_c) \left[ 1 \pm \sqrt{1-\frac{2\tau v_0 (\tau v_0 + \beta a \phi_0)}{\alpha^2 (\phi_0-\phi_c)^2}q^2} \right]$. We also find a pure transverse orientational mode, with the pulsation $\omega_\perp = i \frac{\alpha}{\tau} (\phi_0 - \phi_c)$. We find that the isotropic state is linearly stable below the critical density $\phi_c$. However, it is  unstable against both orientation and compression fluctuations when $\phi_0 > \phi_c$. Again, the fact that compression modes are unstable is consistent with the formation of bands from isotropic phases, which is observed experimentally when $\phi_0 > \phi_c$.\\
 \subsection{Constitutive density-polarization relation in band phases}

At the onset of collective motion, homogeneous states are linearly unstable. The experiments show that large density excitations (bands) steadily propagate in the system. It is difficult to derive analytically the shape of band-density profiles. However, the particle-number conservation provides a relation between the local density and the local polarization field when density excitations  propagate steadily. Looking for propagative solutions of the form $\phi = \phi(x - c_{\textrm{band}}t)$, $\Pi = \Pi(x - c_{\textrm{band}}t) \vec{\hat x}$ and integrating Eq.~(\ref{continuity}) over the transverse direction leads to the relation
\begin{equation}
	\Pi(s) = \frac{c_{\textrm{band}}}{v_0} \left( 1 - \frac{\phi_\infty}{\phi(s)} \right)
\end{equation}
where the integration constant $\phi_\infty$ is the area fraction far away from the band. Note that this expression does not depend on any closure scheme at the hydrodynamic level. The latter relation was used to fit the data, Fig.~3d in the main text. The agreement with our theoretical prediction is a direct proof that the bands are stationary structures as it was also demonstrated in~\cite{Bertin} for the numerical Vicsek model. Band states are genuine phases of  colloidal active matter.\\

 \section{Polar liquid phase}
 \label{polar_liquid}
 
 The closure scheme we followed above (a scaling ansatz for the magnitude of the Fourier modes of $\Psi$) was widely implemented in the previous studies on active matter. We stress that this scheme is not relevant any more for strongly polarized phases. In particular, it results in an unexpected decay of the mean orientation $\Pi_0$ with the density, as shown in Fig.~\ref{fig_transition}B (dashed line). This approximation does not support the observation of stable homogeneous polar liquids, in strong contrast with our experimental findings. This may not be surprising, since the link it provides between  $\vec Q$, $\phi$, and $\vec \Pi$ was devised for weakly-organized phases, an approximation which obviously breaks down in the polar-liquid phase. 
 
 In order to investigate the properties of the polar liquids, we introduce a new closure approximation. When the angular probability distribution $\Psi(\vec r, \theta,t)$ is peaked, the high-order cumulants of the generating function can be neglected. Then a simple assumption, which becomes exact in the limit of perfectly polar order, is to approximate the angular distribution by the wrapped normal distribution, the mean and the variance of which is determined in a self-consistent fashion. The "Gaussian" ansatz imposes the following relation between the angular Fourier components $\hat \Psi_k$ of the distribution function: $\hat \Psi_2 =\vert \hat \Psi_1 \vert^2 \hat \Psi_1 \hat \Psi_1/\hat \Psi_0^3$.  Equivalently, it can also be written as $\vec Q = \Pi^2 \, \vec \Pi \, \vec \Pi - \frac{1}{2} \Pi^4 \, \vec I$, where we have dropped the implicit dependence in $\vec r,t$. With this new closure relation, neglecting higher-order terms in $\frac{a}{H}$, the dynamics of the orientation field is obtained from Eqs.~(\ref{Psi_1})--(\ref{closure1}). Again after some lengthy algebra we obtain:
 \begin{align}
 \label{eq_polar}
	&\tau \partial_t \, \vec \Pi + \tau v_0 \Pi^2 (\vec \Pi \cdot \nabla) \vec \Pi = \left[ \alpha (1 - \Pi^4) \phi - \tau D_r \right] \vec \Pi + \kappa \left[ (1+\Pi^4) \vec I - 2 \Pi^2 \vec \Pi \, \vec \Pi \right] \cdot \vec M * \phi \vec \Pi - \frac{\tau v_0}{2 \phi} (1 - \Pi^4) \nabla \phi \\& \hspace{4cm} \nonumber + \frac{\tau v_0}{\phi} (1- \Pi^2) (\vec \Pi \cdot \nabla \phi) \vec \Pi + \tau v_0 (1- \Pi^2) (\nabla \cdot \vec \Pi) \vec \Pi + \tau v_0 (\Pi^2 \vec I - \vec \Pi \, \vec \Pi) \cdot \nabla(\Pi^2) \\& \hspace{4cm} \nonumber - \frac{1}{2} \beta a \left[ (1+\Pi^4) \vec I - 2 \Pi^2 \vec \Pi \, \vec \Pi \right] \cdot \nabla \phi + \frac{1}{2} \gamma a^2 \left[ (1+\Pi^4) \vec I - 2 \Pi^2 \vec \Pi \, \vec \Pi \right] \cdot \nabla^2 (2 \vec I - \vec M) \cdot \phi \vec \Pi \\& \hspace{4cm} \nonumber+ \mathcal O(\nabla^3)
\end{align}
 where 
 \begin{equation}
 \label{gamma}
 \gamma \equiv \tilde \mu_s \frac{3H}{4\pi a}
 \end{equation}\\

\subsection{Transition to collective motion: Curie-Weiss description}
Even though we built this novel closure scheme to address the properties of the polar liquid, it is worth noting that Eq.~(\ref{eq_polar}) also accounts for the transition to collective motion. Looking again at homogeneous phases, the relation between the average polarization $\Pi_0$ and the average area fraction $\phi_0$  follows from Eq.~(\ref{eq_polar}):
\begin{equation}
\label{order_polar}
	\Pi_0(\phi_0) = \left( 1 - \frac{\phi_c}{\phi_0} \right)^{1/4}
\end{equation}
These variations are  plotted in Fig.~\ref{fig_transition}B (full line). As expected, $\Pi_0$ plateaus to $1$ in the limit of highly concentrated suspensions. More surprisingly, even though the closure scheme is {\em a priori} valid only for ordered phases, the above relation predicts a transition to collective motion at the same critical value $\phi_{\rm c}$ as the one found in Eq.~(\ref{CW_isotropic}).
Therefore, we can reasonably expect Eq.~(\ref{eq_polar}) to be accurate over a wide range of area fractions. However, the $1/4$ scaling at the onset of collective motion is not expected to be valid. Firstly, the usual closure relation, that we used in the previous section, to account for weakly polarized states is not compatible with the Gaussian fluctuation hypothesis. In addition, and more importantly, we know from the experiments that this scaling inferred from a Curie-Weiss approximation  cannot be probed as the system phase separate so that extended ordered bands cruise through an isotropic gaseous phase. This phase separation is reflected by a mere qualitative agreement   between the theory and the measure of the polarization curve shown in Fig. S6B.\\
 
\subsection{Linear stability analysis}
\begin{figure}
\begin{center}
\includegraphics[scale=0.63]{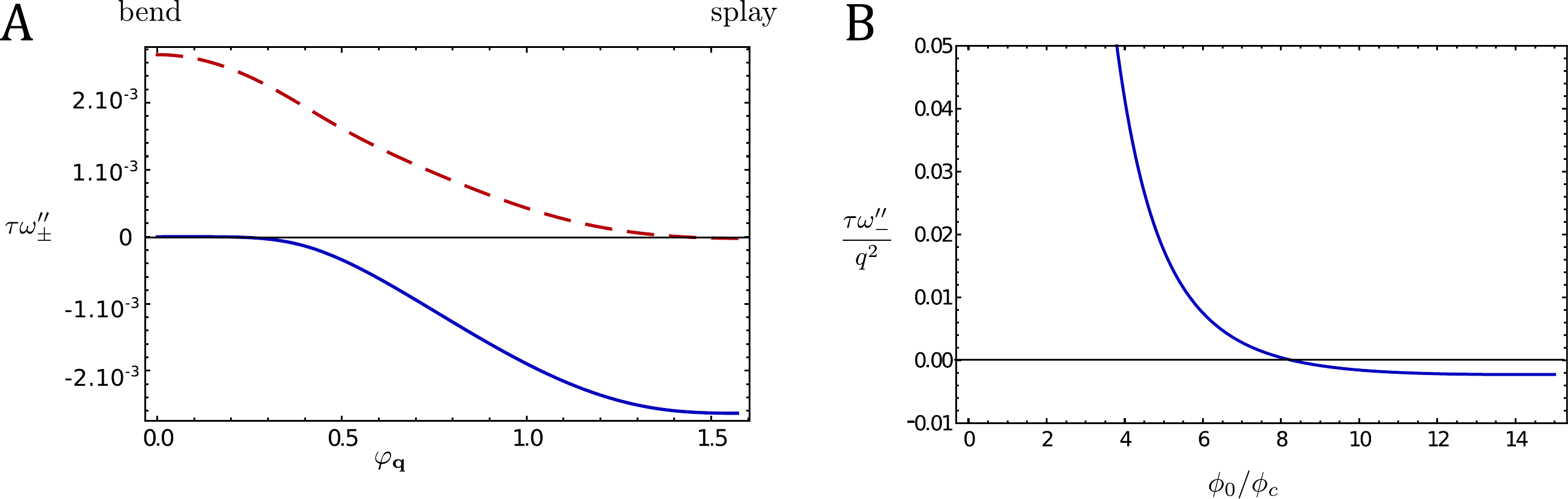}
\end{center}
\caption{\setstretch{1.2}\label{stab2}Stability of strongly polarized states. {\bf A--}~The growth rates $\omega''_\pm$ are plotted as a function of the wave-vector direction $\varphi_\vec q$ Blue line: $\tau \omega''_-$, red dashed line: $\tau \omega''_+$ ($\phi_0 = 10 \phi_c$, $qa = \frac{1}{500}$, $E_0 = 2 E_Q$,  the values of the other parameters are the same as in Fig.~\ref{fig_transition}). {\bf B--}~Magnitude of the pure compression mode $\varphi_{\vec q}=0$ plotted versus the average area fraction of rollers. The repulsive electrostatic interactions result in a restabilization of the compression waves at high area fractions.}
\end{figure}

Following the same approach as in section IV~\ref{linearstabilityweaklypolar}, we now investigate the linear stability of homogeneous polar phases, with respect to spatial fluctuations, for densities $\phi_0 \gg \phi_c$. The mean polarization $\Pi_0$ is given by Eq.~(\ref{order_polar}). We consider plane-wave perturbations of the form $\phi(\vec r,t) = \phi_0 + \delta \hat \phi \, e^{i(\vec q \cdot \vec r - \omega t)}$ and $\vec \Pi(\vec r,t) = \Pi_0 \vec{\hat x} + \delta \vec {\hat \Pi} \, e^{i(\vec q \cdot \vec r - \omega t)}$. The wave-vector direction is defined as $\vec q \equiv q \cos \varphi_\vec q \, \vec{\hat x}+ q \sin \varphi_\vec q \, \vec{\hat y}$. 
Performing a conventional linear stability analysis of Eqs.~(\ref{continuity}) and (\ref{eq_polar}), and restraining ourselves to terms at  zeroth order in $\frac{\phi_0}{\phi_c}$, we find that the dispersions take the form $\omega_\pm(\vec q) \equiv \omega'_\pm(\vec q) + i \, \omega''({\vec q})$. Their explicit expressions are:
\begin{align}
\label{omega1}	&\tau \omega'_\pm = \tau v_0 q \cos \varphi_\vec q \pm \sqrt{\sqrt{F_1^2 + F_2^2}+F_1}\\
\label{omega2}	&\tau \omega''_\pm = F_0 \pm \sqrt{\sqrt{F_1^2 + F_2^2}-F_1}
\end{align}
where $F_0 = \kappa \phi_0 \cos(2 \varphi_\vec q) - \frac{\gamma}{2}a^2 \phi_0 \left(2 - \cos(2 \varphi_\vec q)\right)q^2$, $F_1 = \frac{\beta}{2} a \tau v_0 \phi_0 q^2 \sin^2 \varphi_\vec q - \frac{1}{2}F_0^2$ and $F_2 =\linebreak -2\tau v_0 \phi_0 q \left[ \kappa  + \frac{\gamma}{2} a^2q^2 \right] \sin^2 \varphi_\vec q \cos \varphi_\vec q$. We note again that the above eigenvalues are invariant upon the transformations $\varphi_\vec q \rightarrow - \varphi_\vec q$ and $\varphi_\vec q \rightarrow \varphi_\vec q + \pi$. Without loss of generality, we focus on $\varphi_\vec q \in \left[ 0 , \frac{\pi}{2} \right]$. The growth rates are plotted in Fig.~\ref{stab2}A. Depending on the direction $\varphi_\vec q$, we find positive or negative growth rates, leading to unstable or stable eigenmodes, respectively.

\begin{itemize}

\item {\itshape Splay modes}\quad To further clarify the stabilization/destabilization mechanisms, we expand Eqs.~(\ref{omega1}) and~(\ref{omega2}) in the small wave-vector limit. For $\varphi_{\vec q} >\pi/4$, the eigenvalues  follow simple scaling laws: $\tau \omega'_\pm = \mathcal O(qa)$, $\tau \omega''_+ = \mathcal O(q^2 a^2)$ and $\tau \omega''_- = 2 \kappa \phi_0 \cos(2 \varphi_\vec q) + \mathcal O(q^2 a^2)$. Importantly, we find that the fastest  rate  $\tau \omega''_-$ scales  as $\tau \omega''_- =-2 \kappa \phi_0+ \mathcal O((qa)^2)$. It corresponds to a pure splay mode ($\varphi_{\vec q}=\pi/2$). Since $\omega''_-$ is proportional to $\kappa$ and negative, splay fluctuations are stabilized by the long-range hydrodynamic interactions. We also emphasize that at leading order in $qa$ the relaxation rate $\vert \omega''_- \vert$ does not depend on the wave-vector, i.e. the stabilization of the corresponding splay mode is generic, for the very same reason as the one we discussed in~\cite{Brotto}. This important observation plays a central role in the suppression of giant density fluctuations, as we discuss it in the next section.

\item {\itshape Bending modes}\quad For $\varphi_{\vec q} < \pi/4$, a similar small-$q$ expansion yields $\tau \omega'_\pm = \mathcal O(qa)$, $\tau \omega''_+ = 2 \kappa \phi_0 \cos(2 \varphi_\vec q) + \mathcal O(q^2 a^2)$ and $\tau \omega''_- = \mathcal O(q^2 a^2)$. Fluctuations having the form of bend modes ($\varphi_\vec q = 0$) are exponentially amplified. However, as obviously expected transverse confinement eliminates this instability. More details about  confinement-induced stabilization will be provided  in a forthcoming detailed paper.

\item {\itshape Compression mode}\quad At leading order in $ \left( \frac{\phi_0}{\phi_c} \right)$,  the pure-compression mode corresponding to $\varphi_\vec q = 0$, and $\delta \Pi_x \neq 0$ is marginally stable, i.e. compression fluctuations are merely advected at a velocity $v_0$. Investigating their linear stability requires to expand the  equations of motion up to order $ \left( \frac{\phi_0}{\phi_c} \right)^2$. This yields the following growth rate:
\begin{equation}
	\tau \omega_-'' = \frac{3 (\tau v_0)^2}{64 \alpha \phi_0} \left( \frac{\phi_c}{\phi_0} \right)^2 q^2 - \frac{\beta}{8 \alpha} a \tau v_0 \frac{\phi_c}{\phi_0} \left[ 1 + \frac{\phi_c}{\phi_0} \right] q^2 + o(q^2 a^2)
\end{equation}
It is plotted as a function of $\phi_0$ in Fig.~\ref{stab2}B. The compression mode is unstable below a critical area fraction, and becomes stable at higher densities. To gain more physical insight into this bifurcation mechanism, we propose the following qualitative explanation. Below a critical area fraction ($\sim 10\,\phi_{\rm c}$ within our experimental conditions) the first term of the above expression leads to a positive growth rate. It originates from the alignment interactions, which destabilize the compression fluctuations. When the density locally increases, alignment is enhanced and the local polar order increases accordingly: $\delta \Pi_x > 0$. As a consequence, concentrated regions move coherently as a "rigid body" through a less concentrated background. As a result this "coherent pack "captures even more particles due to the alignment interactions with the particles that collide it. The initial density fluctuations are thus amplified. This is consistent with the weakly polarized state being unstable near the transition to collective motion, as we found in section~\ref{transition}. However, the electrostatic repulsion, proportional to $\beta$, impedes the formation of  concentrated regions. It results in a second term which stabilizes polar liquid phases above a critical density, see Fig.~\ref{stab2}B. Again this prediction is in good agreement with our experimental observations. We find that the band phase evolves into  an homogeneous polar-liquid state as the roller density is sufficiently increased, see Fig.~2 in the main text.
\end{itemize}

To close this section, it is worth noting that hydrodynamics was shown to destroy polar ordering in several models of active suspensions~\cite{Marchetti,Saintillan}. These papers focus on 3D suspensions of microswimers. The flow disturbance induced by the swimmer on the surrounding fluid is modeled as a force dipole singularity (in the far field).  Moreover, the swimmers reorient along the principal direction of the local elongation of the flow, and rotate due to the local vorticity  (Jeffery's orbits). 
The resulting equations of motion for the density and the polarization fields are linearly unstable around a homogeneous polar state. The hydrodynamic dipoles result in a generic suppression of polar ordering (the growth rate of the instability does not depend on the wavelength at long-enougth wavelengths)~\cite{Saintillan}. However, in our system the propulsion mechanism, the rolling of the colloids, strongly involve the lower surface. As they roll, the colloids induce far-field perturbations that have the symmetry of a mirrored rotlet. The magnitude of this singular perturbation to the flow can have a non-zero value because  momentum is continuously exchanged between the fluid and the confining walls.  In addition, the coupling to the local flow field also differs from the one considered for unbounded suspensions. As shown in the previous sections, the rollers  align with the local flow on the bottom-surface. Both the symmetry of the far-field flow, and the local coupling to the fluid flow result in an alignment interaction  between the colloidal rollers. The emergence of polar order from an isotropic population  is a direct consequence of this microscopic polar interaction. Therefore the qualitative consequences of hydrodynamic interactions on the large-scale behavior of active fluids strongly depend on the microscopic mechanism responsible for self-propulsion.\\

\subsection{Density fluctuations}
We now turn to the question of density-density correlations in the polar liquid phase. Most of all the previous theoretical studies on polar active matter have reported the emergence of "giant density fluctuations" in polar liquids. Giant density fluctuations were believed to be a robust and generic feature of active polar liquids~\cite{Marchetti, Toner}. Although our model includes classical alignment interactions, it also contains additional repulsive and dipolar couplings: among them, the long-range hydrodynamic interactions destroy the giant density fluctuations, as we show below. 

To account for the density fluctuations, we employ a fluctuating-hydrodynamic description. We add a conserved white noise term $\nabla \cdot \boldsymbol \xi_\phi$ to Eq.~(\ref{continuity}) and a non-conserved Gaussian noise $\xi_\Pi \vec{\hat y}$ to Eq.~(\ref{eq_polar}), with zero mean and correlations $\left\langle \xi_{\phi_m}(\vec r,t) \xi_{\phi_n}(\vec r',t') \right\rangle \equiv 2 D_\phi \,\delta_{m,n} \delta(\vec r - \vec r') \delta(t-t')$, $\left\langle \xi_\Pi(\vec r,t) \xi_\Pi(\vec r',t') \right\rangle \equiv 2 D_\Pi \, \delta(\vec r - \vec r') \delta(t-t')$, $\left\langle \xi_{\phi_m}(\vec r,t) \xi_\Pi(\vec r',t') \right\rangle \equiv 0$. The correlation function $\left\langle \vert \delta \phi_{\vec q, \omega} \vert^2 \right\rangle$ is calculated in Fourier space within a linear response approximation. The static structure factor for rollers enclosed in a region of area~$\mathcal A$ is defined as:
\begin{equation}
	S(\vec q) \equiv \frac{1}{\pi a^2 \phi_0 \mathcal A} \int_{-\infty}^{+\infty} \frac{\d \omega}{2 \pi} \left\langle \vert \delta \phi_{\vec q, \omega} \vert^2 \right\rangle
\end{equation}
Computing the field amplitude that linearly responds to the noise sources, and averaging over the noise realizations we obtain a rather complex expression for the static structure factor of the polar liquid:
\begin{equation}
	S(\vec q) = \frac{16 \pi^2 \tau}{a^2 \phi_0 \mathcal A} \left[ \left(v_0^2 \phi_0^2 \sin^2 \varphi_\vec q \, D_\Pi + 4 \kappa^2 \phi_0^2 \cos^2(2 \varphi_\vec q) D_\phi \right) q^2 I_1(\vec q) + D_\phi q^2 I_2(\vec q) \right]
\end{equation}
where
\begin{align}
	I_1(\vec q) &= \frac{\vert \omega''_+\vert + \vert \omega''_-\vert}{2 \tau^3 \vert \omega''_+ \omega''_-\vert \left[ (\omega'_+ - \omega'_-)^2 + (\vert \omega''_+\vert + \vert \omega''_-\vert)^2 \right]}\\
	I_2(\vec q) &= \frac{\vert \omega''_+\vert {\omega'_-}^2 + \vert \omega''_-\vert {\omega'_+}^2 + \vert \omega''_+ \omega''_- \vert (\vert \omega''_+\vert + \vert \omega''_-\vert)}{2 \tau \vert \omega''_+ \omega''_-\vert \left[ (\omega'_+ - \omega'_-)^2 + (\vert \omega''_+\vert + \vert \omega''_-\vert)^2 \right]}
\end{align}
where $\omega'_\pm$ and  $\omega''_\pm$ are given by Eqs.~(\ref{omega1}) and (\ref{omega2}). In the small $q$ limit, at leading order we readily find that  the structure factor behaves as $S(q) = \mathcal O((qa)^0)$.  The density fluctuations saturate as  $q \to 0$. In this limit, the structure factor quantifies the large-scale density fluctuations:  $S(q \to 0) = \frac{(\Delta N)^2}{\langle N \rangle}$, where $\langle N \rangle$ is the mean particle number and $(\Delta N)^2$ is the variance of the particle number $N$. It then follows from the saturation of $S$ that the density fluctuations are normal
\begin{equation}
\label{number_fluctuations}
	\frac{\Delta N}{\langle N \rangle} \sim \frac{1}{\sqrt{\langle N \rangle}}
\end{equation}
In a suspension of Quincke rolling particles, the large-scale number fluctuations follow the same scaling law as in equilibrium systems. Contrary to most of active systems, there are \textit{no} giant density fluctuations. This unusual behavior stems from the generic stabilization of splay disturbances $\omega''_-(q) = \mathcal O(1)$ by the long-range dipolar hydrodynamic interactions, which decay like $r^{-2}$ in two dimensions~\cite{Brotto}.

Importantly, the previous asymptotic expansions are valid in the limit of small wave vectors $qH \ll 1$ (we recall that $H$ is the channel height). The long-range dipolar interactions, that govern the density fluctuations at large scales, are  subdominant at  distances smaller than $H$ (see section~\ref{interactions}). As a consequence, deviations from the above prediction are expected below a crossover length of order $H$. In the case of splay fluctuations, i.e. $\varphi_\vec q = \pi/2$, the small-$q$ expansion yields a simple analytic expression for the structure factor at all $q$s:
\begin{equation}
	S(q \, \vec{\hat y}) = S_0 \left( 1 + \zeta q^2 \right) + \mathcal O(q^3 a^3)
\end{equation}
where $S_0 = \frac{8 \pi^2 v_0}{\mathcal A \beta a^3} \left( D_\Pi + \frac{4 \kappa^2}{v_0^2} D_\phi \right)$ and $\zeta = \frac{\beta a \tau v_0}{4 \kappa^2 \phi_0^2} \left( \phi_0 + \frac{2\kappa D_\phi}{4 \kappa^2 D_\phi + v_0^2 D_\Pi} \right)- \frac{3 \gamma}{2 \kappa} a^2$. The structure factor therefore deviates from its large-scale behavior when $q \gtrsim \vert \zeta \vert^{-1/2}$. From the expressions of the coefficients $\beta$, $\kappa$, $\gamma$ that we deduced from the microscopic model of the roller propulsion and interaction mechanisms, Eqs.~(\ref{beta}), (\ref{kappa}) and~(\ref{gamma}), we indeed find that $S({\vec q})$ deviates from a constant value and decays algebraically as $q$ exceeds $H^{-1}$.

A direct comparison with the experimental data is difficult to obtain in stadium-shape potentials due to the small size of our ITO coated slides which prevent the exploration of very small $q$ modes. However, we were able to quantify the particle-number fluctuations in real space. The data unambiguously show that the number fluctuations are normal  $\Delta N \sim \sqrt{\langle N \rangle}$, as expected from Eq.~(\ref{number_fluctuations}). 

As a final remark we stress that the hydrodynamic interactions between the rollers play a curtail role in determining the large-scale behavior of the populations of active colloids. They provide the very mechanism that allows the rollers to sense the orientation of their neighbors and to promote local alignment of their velocities. This microscopic alignment interaction was shown to yield very large-scale orientational order.  Furthermore the hydrodynamic couplings also stabilize the density fluctuations of these unique active polar liquids by suppressing the splay fluctuations of the local polarization field that are generically responsible for the usual giant density fluctuations of active matter~\cite{Schaller}. \\


\titlespacing*{\section}{0pt}{0ex plus 0ex minus 0ex}{0ex plus 0ex}
\begin{center}
\rule{0.5\linewidth}{1pt}
\end{center}
\singlespacing
\begin{small}

\end{small}


\begin{thebibliography}{21}

\bibitem{Melcher} Melcher, J. R. \& Taylor, G. I. Electrohydrodynamics: a review of the role of interfacial shear stresses. {\itshape Annu. Rev. Fluid. Mech.} {\bf 1}, 111--146 (1969).

\bibitem{Pannacci} Pannacci, N., Lobry, L. \& Lemaire, E. How insulating particles increase the conductivity of a suspension. {\itshape Phys. Rev. Lett.} \textbf{99}, 094503 (2007).

\bibitem{Goldman1} Goldman, A. J., Cox, R. G. \& Brenner, H. Slow viscous motion of a sphere parallel to a plane wall -- I Motion through a quiescent fluid. {\itshape Chem. Eng. Sci.} \textbf{22}, 637 (1967).

\bibitem{Goldman2} Goldman, A. J., Cox, R. G. \& Brenner, H. Slow viscous motion of a sphere parallel to a plane wall -- II Couette flow. {\itshape Chem. Eng. Sci.} \textbf{22}, 653 (1967).

\bibitem{O'Neil} O’Neill, M. E., \& Stewartson, K. On the slow motion of a sphere parallel to a nearby plane wall {\itshape J. Fluid Mech.} \textbf{27}, 705 (2006).

\bibitem{Liu} Liu, Q. \& Prosperetti, A. Wall effects on a rotating sphere. {\itshape J. Fluid Mech.} \textbf{657}, 1 (2010).

\bibitem{Das}
Das, D. \& Saintillan, D. Electrohydrodynamic interaction of spherical particles under Quincke rotation. {\itshape Phys. Rev.~E} \textbf{87}, 043014 (2013).

\bibitem{Blake} Blake, J.R . \& Chwang, A. T. Fundamental singularities of viscous flow. {\itshape J. Eng. Math.} \textbf{8}, 23 (1974).

\bibitem{Hackborn} Hackborn, W. W. Asymmetric Stokes flow between parallel planes due to a rotlet. {\itshape J. Fluid Mech.} \textbf{218}, 531 (2006).

\bibitem{Farrell} Farrell, F. D. C., Marchetti, M. C., Marenduzzo, D. \& Tailleur, J. Pattern formation in self-propelled particles with density-dependent motility. {\itshape Phys. Rev. Lett.}  {\bf 108}, 248101 (2012).

\bibitem{Risken} H.~Risken, \textit{The Fokker-Planck Equation: Methods of Solution and Applications}, (Springer Verlag, Berlin, 1996).

\bibitem{Menzel} Menzel, A. Collective motion of binary self-propelled particle mixtures. {\itshape Phys. Rev. E} \textbf{85}, 021912 (2012).

\bibitem{Desreumaux} Desreumaux, N., Caussin, J.-B., Jeanneret, R., Lauga, E. \& Bartolo, D. Hydrodynamic fluctuations in confined emulsions. Preprint at http://arxiv.org/abs/1301.5549 (2013).

\bibitem{Hinch} Hinch, E. J. \& Leal, L.G. Constitutive equations in suspension mechanics. Part 2. Approximate forms for a suspension of rigid particles affected by Brownian rotations. {\itshape J. Fluid Mech.} \textbf{76}, 187 (2006).

\bibitem{Baskaran} Baskaran, A. \& Marchetti, M. C. Hydrodynamics of self-propelled hard rods. {\itshape Phys. Rev. E} \textbf{77}, 011920 (2008).

\bibitem{Woodhouse} Woodhouse, F.G. \& Goldstein, R. E. Spontaneous circulation of confined active suspensions. {\itshape Phys. Rev. Lett.} \textbf{109}, 168105 (2012).

\bibitem{Bertin} Bertin, E., Droz, M. \& Gr\'egoire, G. Hydrodynamic equations for self-propelled particles: microscopic derivation and stability analysis. {\itshape J. Phys. A} {\bf 42}, 445001 (2009).

\bibitem{Peshkov} Peshkov, A., Aranson, I., Bertin, E., Chaté, H. \& Ginelli, F. Nonlinear field equations for aligning self-propelled rods. {\itshape Phys. Rev. Lett.} \textbf{109}, 268701 (2012).

\bibitem{Brotto} Brotto, T., Caussin, J.-B., Lauga, E. \& Bartolo, D. Hydrodynamics of confined active fluids. {\itshape Phys. Rev. Lett.}  {\bf 110}, 038101 (2013).

\bibitem{Marchetti} Marchetti, M.C {\itshape et al}. Soft active matter. Preprint at http://arxiv.org/abs/1207.2929 (2012).

\bibitem{Saintillan} Saintillan, D. \& Shelley, M. J. Instabilities and pattern formation in active particle suspensions: kinetic theory and continuum simulations. {\itshape Phys. Rev. Lett.}  {\bf 100}, 178103 (2008).

\bibitem{Toner} Toner, J., Tu, Y. \& Ramaswamy, S. Hydrodynamics and phases of flocks. {\itshape Ann. Phys.} \textbf{318}, 170 (2005).

\bibitem{Schaller}
Schaller, V. \& Bausch, A. R. Topological defects and density fluctuations in collectively moving systems. {\itshape Proc. Natl. Acad. Sci. USA}  {\bf 110}, 4488--4493 (2013).

\end{thebibliography}
\end{document}